\documentclass[pdflatex,sn-basic]{sn-jnl}

\usepackage{graphicx}%
\usepackage{multirow}%
\usepackage{amsmath,amssymb,amsfonts}%
\usepackage{amsthm}%
\usepackage{mathrsfs}%
\usepackage[title]{appendix}%
\usepackage{xcolor}%
\usepackage{textcomp}%
\usepackage{manyfoot}%
\usepackage{booktabs}%
\usepackage{algorithm}%
\usepackage{algorithmicx}%
\usepackage{algpseudocode}%
\usepackage{listings}%

\usepackage{xfrac}
\usepackage{tikz}
\usepackage{bm,upgreek}
\usepackage{subcaption}
\usepackage{array}
\usetikzlibrary{shapes.geometric, arrows}
\usepackage{ifthen}
\usepackage{wrapfig}
\usepackage{lscape}
\usepackage{rotating}
\usepackage{epstopdf}
\usepackage{tabularx}
\usepackage{anyfontsize}
\usepackage{soul}

\theoremstyle{thmstyleone}%
\theoremstyle{thmstyletwo}%

\theoremstyle{thmstylethree}%

\begin{document}

\title[Slip-tendency analysis of a potential CO\textsubscript{2} storage site in the central German North Sea]{Slip-tendency analysis of a potential CO\textsubscript{2} storage site in the central German North Sea}


\author*[1]{\fnm{Hendrawan D.B.} \sur{Aji}}\email{hendrawan.aji@ifg.uni-kiel.de}

\author[2]{\fnm{Fabian} \sur{Jähne-Klingberg}}\email{fabian.jaehne-klingberg@bgr.de}

\author[2]{\fnm{Heidrun L.} \sur{Stück}}\email{heidrunlouise.stueck@bgr.de}

\author[1]{\fnm{Frank} \sur{Wuttke}}\email{frank.wuttke@ifg.uni-kiel.de}

\author[3]{\fnm{Petia} \sur{Dineva}}\email{petia@imbm.bas.bg}

\affil*[1]{\orgdiv{Institute of Geosciences, Chair of Geomechanics and Geotechnics}, \orgname{Kiel University}, \orgaddress{\street{Ludewig-Meyn-Straße 10}, \city{Kiel}, \postcode{24118}, \country{Germany}}}

\affil[2]{\orgname{Federal Institute for Geosciences and Natural Resources (BGR)}, \orgaddress{\street{Stilleweg 2}, \city{Hannover}, \postcode{30655}, \country{Germany}}}

\affil[3]{\orgdiv{Institute of Mechanics}, \orgname{Bulgarian Academy of Sciences}, \orgaddress{\street{Acad. G. Bontchev St., bl. 4}, \city{Sofia}, \postcode{1113}, \country{Bulgaria}}}


\abstract{Underground CO\textsubscript{2} storage operations may lead to induced seismicity, which can be partially related to nearby affected faults. Considering the storage potential in the Middle Buntsandstein sandstone formation on the West Schleswig Block within the German North Sea, we investigate the reactivation and seismicity potential of existing faults in the vicinity of a selected study area. For this, we develop a 3D fault data set based on roughly 80 2D seismic lines, combining 60 years of exploration history. Contextual details regarding the geological features and the lithological description of the area are presented. Using the fault data set and the present-day stress tensors derived through spatial interpolation of the 3D numerical stress model of Germany, a slip-tendency analysis is performed to assess the susceptibility of existing faults to reactivation and reveal critical focal mechanisms, which are important for further wave propagation simulation. Furthermore, we extend the procedure to estimate the sustainable pore pressure window and the potential moment magnitude in the event of a reactivation. The results show that the NW/SE and NNW/SSE-oriented faults exhibit a higher slip tendency. Among these faults, those with a combination of above-average slip tendency and potential moment magnitude are identified and discussed.}




\maketitle
\section{Introduction}
\label{sec1}

It is well-known that earthquakes are unpredictable phenomena caused by naturally shifting stresses in Earth's crust. However, a range of human activity can also induce earthquakes. Many industrial activities today have an influence on the stress-strain state or on the pore pressure within the Earth’s crust that can induce or trigger earthquakes. This happens mainly in geothermal and extraction fields or mining regions. The recent increase of induced seismicity has been linked to subsurface activities related to the followings: \textit{(a)} groundwater extraction (\cite{Amos2014}); \textit{(b)} wastewater disposal (\cite{Keranen2014,Frohlich2016}; \textit{(c)} oil and gas production (\cite{Ellsworth2013, Hough2014, Clarke2014}); \textit{(d)} geothermal energy production (\cite{Giardini2009, Diehl2017}); \textit{(e)} geological carbon dioxide (CO\textsubscript{2}) sequestration (\cite{White2016, STORK2018, GOERTZALLMANN2024}); or \textit{(f)} surface and underground mining (\cite{GIBOWICZ2009}). Literature reviews covering an extensive range of topics including mining-induced seismicity, geotechnical applications, monitoring of petroleum reservoirs, fluid injections in geothermal areas, seismicity associated with water reservoirs, and underground nuclear explosions are presented in \cite{Guha2001, Trifu2002, Ellsworth2013, NAP13355, Grigoli2017, Keranen2018}, among others, while those focused on the risks of induced seismicity related CO\textsubscript{2} storage can be found in \cite{White2016, VERDON2016, CHENG2023}.

Carbon capture and storage (CCS) is seen as a way of reducing carbon emissions into the atmosphere in the near future and storing currently unavoidable emissions deep underground (\cite{VERDON2011,STORK2018,Luu2022}). Numerous studies have shown that induced seismicity risks related to CO\textsubscript{2} storage operations can be minimized through modeling (i.e., site characterization followed by statistical forecasting or numerical simulation), monitoring, and control (i.e., traffic light system) (\cite{IPCC2005, VERDON2016, CHENG2023}). To date, Cogdell field (Texas, USA) is the only known CO\textsubscript{2}-injection field where earthquakes with magnitude $>$1 have been reported (\cite{Gan2013, White2016, CHENG2023}). In Germany, potential CO\textsubscript{2} storage operations in the German North Sea have gained significantly increasing support and interest since 2022, motivated by the German Government's goal to achieve climate neutrality by 2045 (\cite{Wallman2023}).

\begin{figure*}[t!]
\centering
{\includegraphics[trim=0 0 0 0, clip, width=\linewidth]{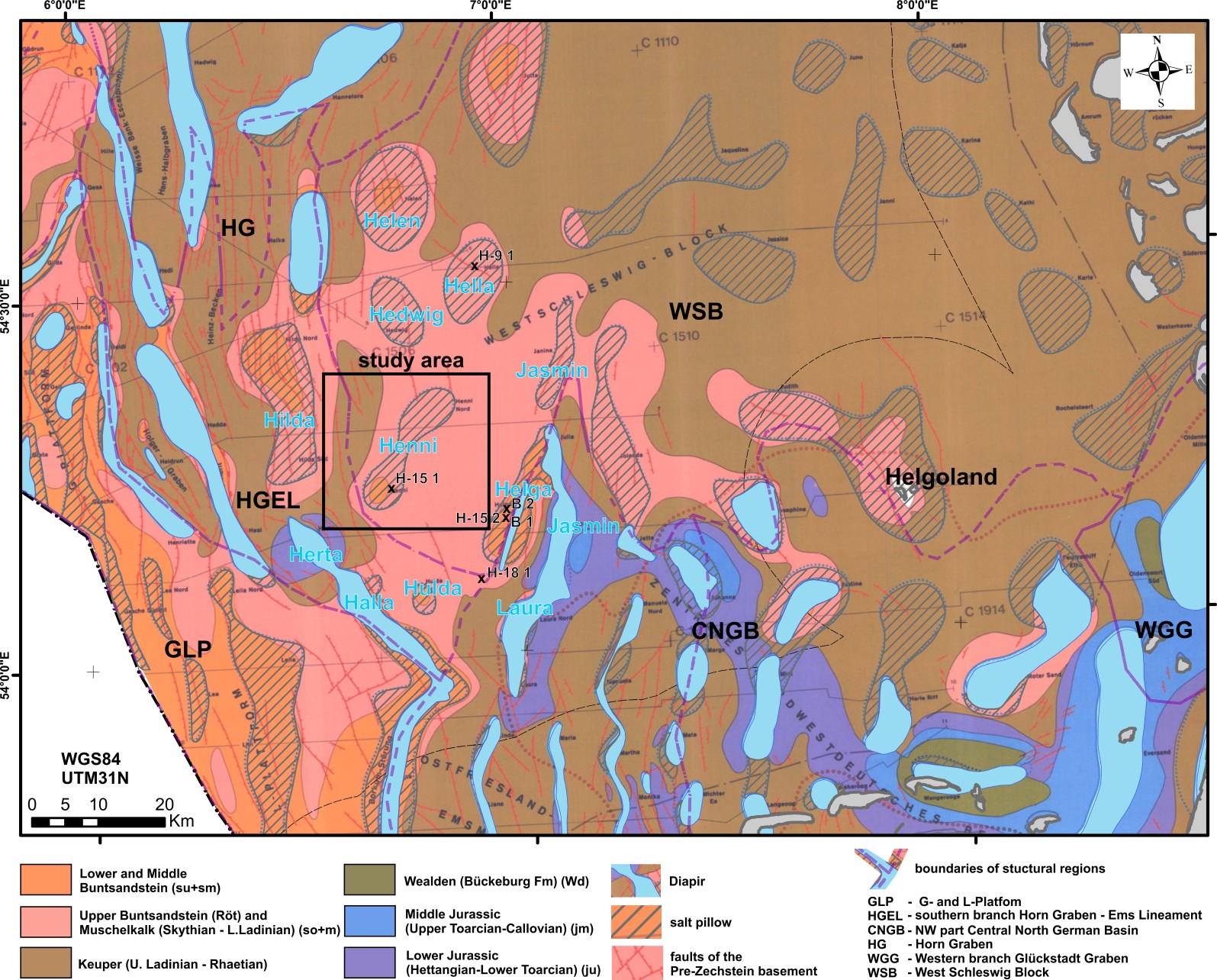}}
\caption{Geological overview map of the central German North Sea and coastal islands with the study region (Henni salt pillow) on the southwestern of the West Schleswig Block. The units in the footwall of the Lower Cretaceous to Latest Upper Jurassic unconformity (barrier unit) according to \cite{Baldschuhn1994} are shown. Salt diapirs and salt pillows, as well as the faults of the Pre-Zechstein basement are also highlighted. Larger structural regions are separated by a dashed line.}
\label{fig:MainMap}
\end{figure*}

Recent investigations on potential storage sites in the Middle Buntsandstein on the West Schleswig Block in the German North Sea (see Fig. \ref{fig:MainMap}) pointed to preferable units due to its geological and petrological characteristic (\cite{FUHRMANN2024, BENSE2017}). The German North Sea itself is a seismically quiet region (\cite{Grunthal2003,Leydecker2011,HEIDBACH2018}), and, thus, a focal mechanism identification through inversion is troublesome to carry out. To overcome this gap, we perform a slip-tendency analysis on the existing faults in the vicinity of a potential storage structure, which allows the evaluation of their susceptibility to reactivation considering the \textit{in-situ} stress field, the faults' geometry, and the effect of non-uniform stress distribution. The slip-tendency analysis has been proven useful for assessing the seismic risks in areas with high natural seismicity (\cite{Morris1996, COLLETTINI2007}), a region with geothermal exploration potential (\cite{MOECK2009, Roeckel2022}), and a wastewater injection field (\cite{Vadacca2021}). The study by \cite{MOECK2009} also suggested that the results from such an analysis are well correlated with low-magnitude seismicity recorded in a stimulated area. In addition to identifying the critical faults and focal mechanisms, we extend the procedure to derive the sustainable pore pressure window and estimate the potential moment magnitude in the event of a slip. The former is a determining factor in the design and optimization of an injection strategy, while the latter is, in conjunction with the focal mechanism, a substantial input for wave propagation simulations.

The present work is organized as follows. First, Section 2 describes the study area and its geological and lithological characteristics. Section 3 discusses the methods used in this study. The result of the slip-tendency analysis is presented and discussed in Section 4. Finally, the paper ends in Section 5 with a list of conclusions and comments for further work.

\section{Study area}\label{sec2}

The area of focus is located in the southwestern part of the West Schleswig Block in the German North Sea. We adopt the result of a recent study by \cite{FUHRMANN2024}, which demonstrated several potential storage structures in the area. We are particularly interested in the structure designated as \textbf{CNS\_24} in their paper (see Fig. \ref{fig:MainMap}), which is bulging in the area of a salt pillow in a hydraulically open to semi-open system. The potential reservoir is located in the base of the Middle Buntsandstein, at a depth of $\approx$1.6 km, and favored due to its large potential static capacity, minimum lateral confinement, top seal $>$20 m, and an economically feasible depth \citep{FUHRMANN2024}. 

In the following, we will first provide a general overview of the geological setting of the study area based on existing knowledge. Next, we will present the database available for this study, followed by the interpretation derived from this data to give a detailed description of the storage structure, focusing on faults. Finally, potential uncertainties underlying this interpretation will be presented.

\subsection{Overview of the geological setting}

The southeastern German North Sea (Fig. \ref{fig:MainMap}) represents the transition from the central North German Basin (CNGB) and the Mesozoic subsidence centre in the area of the Glückstadt Graben (WGG) to the platform areas bordering to the north (the West Schleswig Block/WSB) and northwest (the G- and L-Platform/GLP). These in turn are dissected by the Horn Graben (HG), another Mesozoic graben similar to the WGG. The Mesozoic platforms are currently the main focus of interest in the search for suitable structures for CO\textsubscript{2} storage. The sedimentary platforms, such as the WSB, are characterised by uniform, steadily progressive sedimentation with low sedimentation rates over long periods of time and, in comparison to other regions, show minor faulting of the Mesozoic to Cenozoic overburden, e.g., see \cite{Kockel1995}. In particular, during the Mesozoic basin development, deposition and erosion in the WSB area was largely uniform over large areas. Salt flow triggered by rifting in the area of the HG, WGG, but also by increased subsidence of the neighbouring CNGB to the south, may have led to the formation of larger, more strongly bulging salt pillows along the borders of the WSB to neighbouring structural units during the Middle Triassic. In general, salt pillows in the North German Basin are considered to be particularly suitable trap structures for the potential storage of fluids in the Middle Buntsandstein saline aquifer (e.g., see \cite{Jaehne2014,FUHRMANN2024}).

\begin{figure}
\centering
{\includegraphics[trim=0 0 0 0, clip, width=0.6\linewidth]{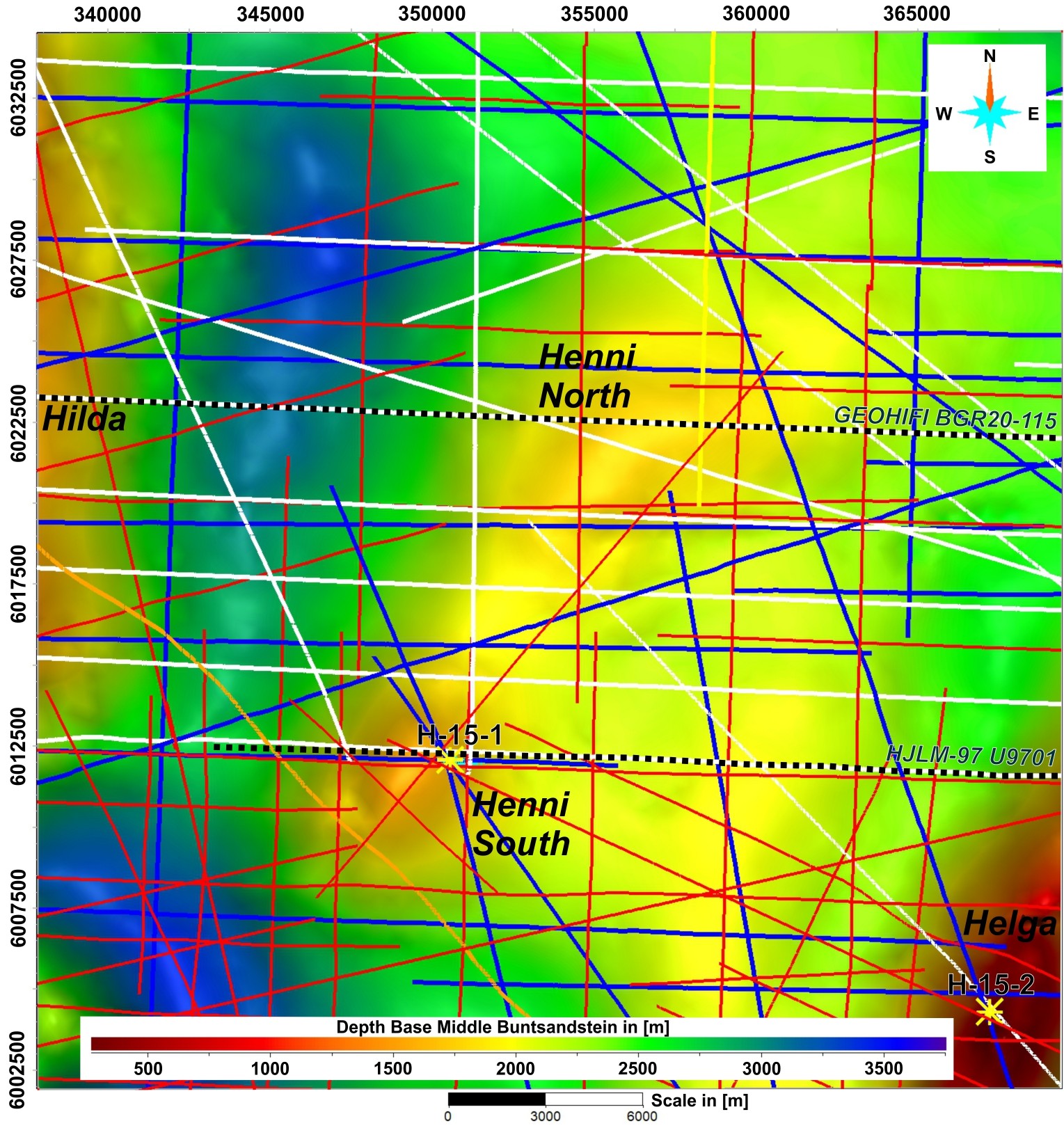}}
\caption{Overview of the seismic and well data surrounding Henni salt pillow. The base of the Middle Buntsandstein reservoir is shown in the background. The 2D-seismic surveys are marked in different colors according to their age, resolution with depth, and processing status: (white) mid-depth high-resolution 2D survey GEOHIFI acquired by BGR in 2020 \citep{Ehrhardt2021}; (yellow) shallow 2D seismic profile BGR Aurelia 2003; (blue) digitally available 2D profiles from the 1980s/1990s; (red) processed vintage data from the 1970s\textendash1980s; (orange) processed seismic line from the mid-1960s. Two interpreted seismic profiles (Fig. \ref{fig:CrossSeismic}) are highlighted by dotted lines. Coordinate system ETRS 89 UTM 32N.}
\label{fig:Henni2DSeismicLines}
\end{figure}

\subsection{Database}

The Henni salt pillow is covered or at least partially intersected by almost 80 2D seismic lines from 17 seismic surveys with a nearly 60-year exploration history, processed with different workflows and parameters and overall with different levels of processing (Fig. \ref{fig:Henni2DSeismicLines}). The average distance between high-detailed digitally available lines from the late 1980s to the 1990s and lines of the GEOHIFI survey recorded in 2020 is around 3 km. However, on the southern flank of Henni South, the distance between digitally available data is almost 6 km. To perform a more detailed analysis of fault patterns in this area, several seismic lines from vintage surveys (1960s\textendash1980s) were additionally processed and interpreted. For the southern flank of Henni, the integration of vintage data results in some significantly different horizon geometries with deviations of up to 150 ms (roughly between 190 to 300 m) compared to modeling without vintage data. However, WE or even SE-striking faults in the area of Henni South become visible only through the integration of vintage data. Of all things, a NW-SE trending line from the 1960s (profile NORD 65-237) provides the greatest insight into the southern flank of Henni.

In addition to the seismic data, Henni structure has been drilled down to the level of the Lower Buntsandstein by borehole H 15-1. Further deep wells for correlating the seismic data are located in the top and on the flanks of the Helga structure to the east (B 1, B 2, H 15-2, H 18-1) or even further away in the top of the Hella salt pillow to the north (H-9-1). Based on work on velocity modeling from previous projects (\cite{Bense2022,Gross1986,Jaritz1991}), the interpretation results and the faults modeled in seismic two-way travel time (TWT) runtime were transferred to the depth domain for further analysis of fault mechanics.

For the multi-survey interpretation, a seismic-seismic tie was performed. However, this was not always clearly possible due to the different data processing of the individual surveys. Moreover, with older vintage data, it is sometimes no longer possible to localise the shot points exactly, which leads to positional inaccuracies.

\begin{figure*}
\centering
{\includegraphics[trim=0 0 0 0, clip, width=\linewidth]{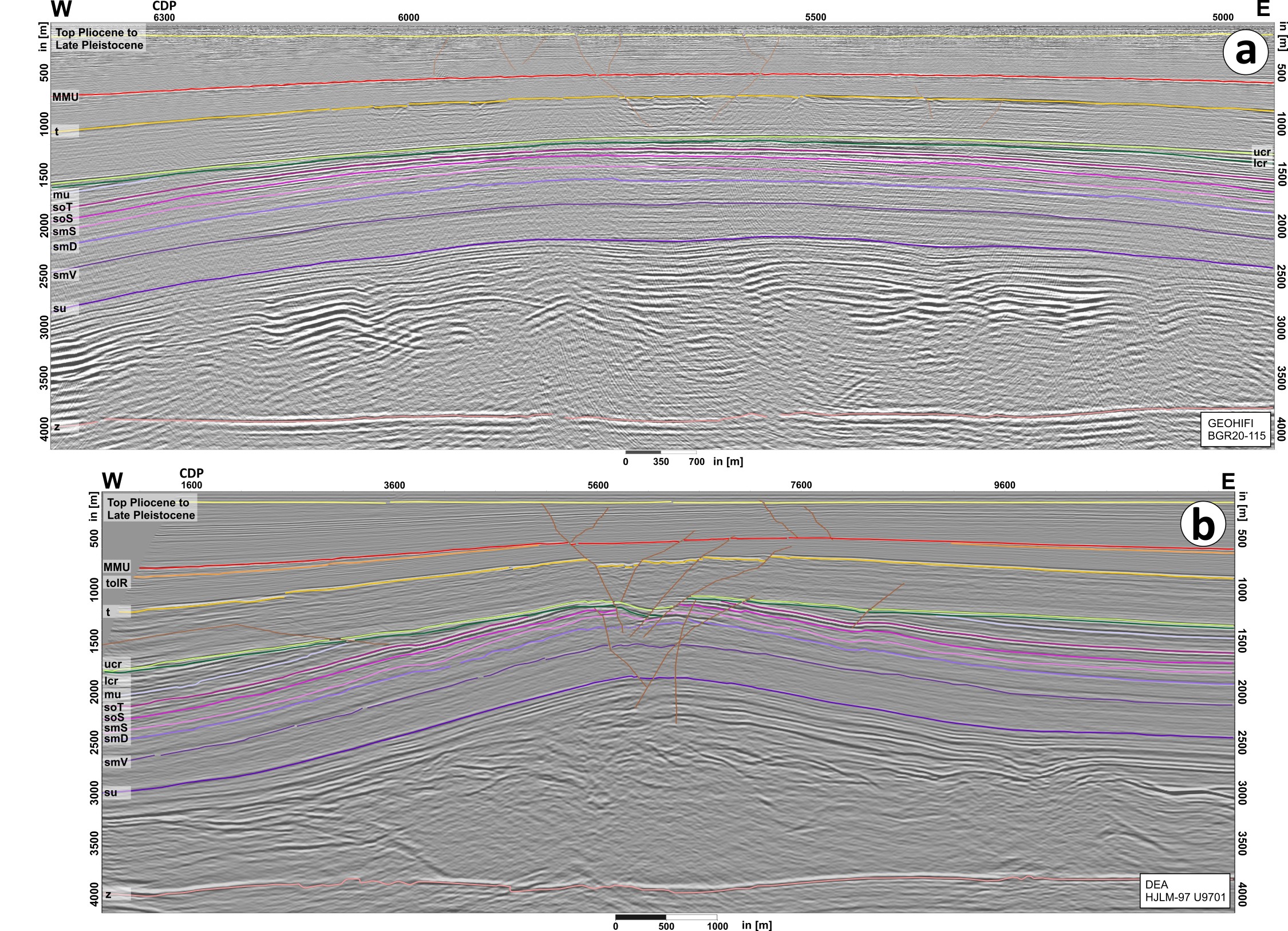}}
\caption{Examplary seismic interpretations for the northern (\textbf{a}: GEOHIFI BGR20-115) and southern (\textbf{b}: DEA HJLM-97 U9701) part of the Henni salt pillow based on approximately EW trending depth-converted seismic lines. The location of each profile is highlighted in Figs. \ref{fig:Henni2DSeismicLines} and \ref{fig:HenniFaultTopview}.}
\label{fig:CrossSeismic}
\end{figure*}

\begin{figure*}
\centering
{\includegraphics[trim=0 0 0 0, clip, width=\linewidth]{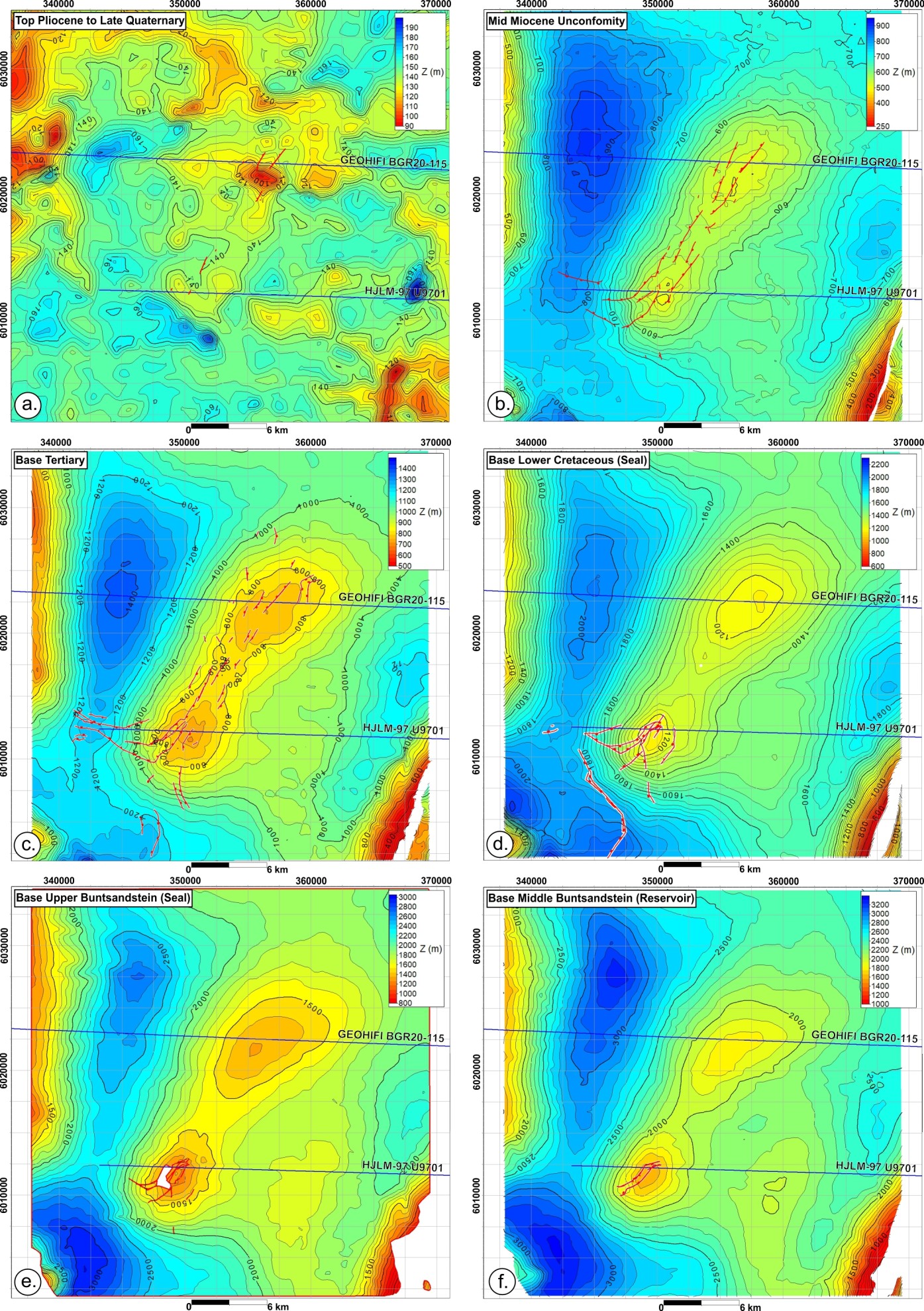}}
\caption{Results of geological mapping and modelling in the area of the Henni salt pillow. The data used for this are shown in Fig. \ref{fig:Henni2DSeismicLines}. The horizons illustrated here show the bases of \textbf{(a\textendash c)} the Tertiary to Quaternary evolutionary history; (\textbf{d} and \textbf{e}) the Upper Buntsandstein and the Lower Cretaceous, which are the barrier formations in the reservoir caprock; and \textbf{(f)} the Middle Buntsandstein, which is the structure of the reservoir under investigation.}
\label{fig:HenniFaultTopview}
\end{figure*}

\subsection{Geological history and structural inventory}\label{sec2 inventory}

Fig. \ref{fig:MainMap} highlights the location of the Henni salt pillow in an area of the WSB that is influenced by pre-Cretaceous uplift and increased erosion down to the level of the Upper Buntsandstein (e.g. \cite{Kockel1995}). No Upper Muschelkalk to Middle Jurassic sediments are preserved on the Henni structure (Fig. \ref{fig:CrossSeismic}), indicating a Pre-Cretaceous uplift and erosion. According to well H 15-1 (Fig. \ref{fig:MechProps}), the stratigraphic record only begins again with the Tithonian in the hanging wall of the Upper Jurassic to Lower Cretaceous unconformity (LCU). The sequences in the hanging wall of the LCU are relatively complete except for smaller erosional gaps. However, variations in the thickness of the individual formations can also be observed for these sedimentary sequences (Figs. \ref{fig:CrossSeismic} and \ref{fig:MechProps}), which is related to a renewed uplift of the Henni salt pillow that lasted from the Upper Cretaceous to the Miocene. This renewed halotectonic development phase of Henni is contemporaneous with the development and formation of Tertiary crestal faults in the top of a number of salt structures in the central German North Sea (see e.g., \cite{Bruekner2005}).

A closer look at the structure (Fig. \ref{fig:HenniFaultTopview}) reveals two structure maxima: Henni North and Henni South. The latter is significantly more bulged than the former and also shows a more pronounced angular unconformity at the LCU (Fig. \ref{fig:CrossSeismic}). The angular unconformity at the base of the Lower Cretaceous and the non-uniform erosion of Triassic sequences across the structure is also a clear indication of a pre-Cretaceous development of the Henni structure. Therefore, a formation of the pillow would be probable for the later Middle Triassic to the Middle Jurassic, whereby a comparison with other salt structures of the German North Sea with more complete sedimentary sequences suggests a formation in the late Middle Triassic to Upper Triassic (e.g., \cite{Kockel1995}). Another striking feature, particularly at Henni South, is that due to the incision of the pre-Cretaceous erosion down to the Upper Buntsandstein, the Roet salinar was dissolved by subrosion over several kilometres from the apex of the structure (Fig. \ref{fig:CrossSeismic}b). It is possible that the influence of the subrosion of the salinar can also be seen in the thickness and facies of the overlying Lower Cretaceous. However, due to the low well density and the lack of 3D seismic data, the nature of the Lower Cretaceous barrier cannot be conclusively assessed for the time being.

As indicated, Henni South shows a stronger tectonic overprint than Henni North, in the course of more pronounced halotectonics both pre-Cretaceous and again from the Upper Cretaceous onwards. In Fig. \ref{fig:CrossSeismic}b, a tree-shaped network of faults can be interpreted, which can be traced from the base of the Buntsandstein to the youngest Cenozoic units (see also Fig. \ref{fig:HenniFaultTopview}). This network of faults is the result of the formation of crestal faults both during the pre-Cretaceous structural development and then during the renewed pillow uplift in the Upper Cretaceous up to the Miocene and the associated partial reactivation of pre-cursors. This pattern does not appear at Henni North (Figs. \ref{fig:CrossSeismic}a \& \ref{fig:HenniFaultTopview}). With increasing sedimentary burial in the course of the formation of the Eridanos delta (e.g., \cite{Thoele2014}) in the hanging wall of the Middle Miocene unconformity, the salt pillow development gradually came to a standstill and thus also lead to a decrease in fault activity. The number of faults and their offsets steadily decreased until the late Quaternary (Figs. \ref{fig:CrossSeismic} \& \ref{fig:HenniFaultTopview}a).

The pre-Zechstein basement beneath Henni only shows faults with small offsets (a few tens of metres), which do not indicate a direct connection to the faults in the Mesozoic to Cenozoic cover (Fig. \ref{fig:CrossSeismic}). The largest fault offsets are interpreted in the top of Henni South with vertical offsets of 100\textendash200 m at the Lower Cretaceous to Upper Cretaceous level. The vertical offsets in the Lower Buntsandstein are significantly smaller. The Upper Buntsandstein at Henni South still shows vertical offsets of $\approx$50 metres. For the reservoir unit of the Middle Buntsandstein, mainly vertical offsets significantly $<$50 m were detected. The offsets at the base of the Tertiary are usually between 50 and 75 metres. Depending on the seismic line, the vertical offsets at the base of the Mid Miocene Unconformity (MMU) and in its hanging wall are barely detectable and usually $<$25 m. The vertical offsets in the area of the uppermost mapped horizon (Fig. \ref{fig:HenniFaultTopview}a), which roughly corresponds to the latest Quaternary, cannot be clearly identified. The faults that reach this structural level probably only show offsets of a few metres. In general, however, it can be stated that the offsets in the top of Henni North (Fig. \ref{fig:CrossSeismic}a) are significantly smaller in total than in the top of Henni South (Fig. \ref{fig:CrossSeismic}b). For the entire structure, the majority of faults are normal faults, but in transitional areas between individual faults and on faults that significantly change their strike, an oblique sense of displacement cannot be ruled out. Reverse faulting or transpressive faulting was not mapped and, if so, can only be assumed to be local and subordinate, but are generally possible in the crestal areas of diapirs and salt pillows \citep{Jackson2017}.

Fig. \ref{fig:HenniFaultTopview} provides an overview of the fault patterns at different structural levels. It can be seen that the faults in the Triassic in the footwall of the LCU are limited exclusively to the top of Henni South. Parts of the Upper Buntsandstein (seal) and the Muschelkalk are also eroded there (Fig. \ref{fig:HenniFaultTopview}e). Especially in the sequences from the Lower Cretaceous to the base of the MMU, along the western and southern flanks of Henni South, radial aligned grabens and faults with W\textendash E and approx. N\textendash S orientation are formed. This set of faults appears to have had no influence on the Middle Buntsandstein reservoir, but in some cases clearly offset the Lower Cretaceous barrier unit. In Fig. \ref{fig:CrossSeismic}b, one of these faults is unfavourably transected (comparison with Fig. \ref{fig:HenniFaultTopview}) at a narrow angle, but appears to have a possible detachment in the Roet (Upper Buntsandstein). Within the Upper Cretaceous, a NNE to NE trending fault set seems to have developed across the entire Henni structure. This fault set can also be observed in the Palaeogene (Fig. \ref{fig:HenniFaultTopview}c) and Miocene (Fig. \ref{fig:HenniFaultTopview}b) units. The faults in the top of Henni North do not appear to be directly linked to those of Henni South, but are diffusely interlocked between the two maxima of the Henni salt structure. Possible linkages of the faults in the top of the Henni North to those of the Henni South cannot be reliably denied with the current data situation.

With regard to potential fluid migration pathways, the following can be noted. The interpretation of the southern sub-trap (Fig. \ref{fig:CrossSeismic}b) points to a strong segmentation at the apex of the structure and fault networks that could represent potential fluid pathways from the reservoir to near the surface. In contrast, faults in the northern part of the salt pillow (Fig. \ref{fig:CrossSeismic}a) are mapped exclusively in the Cenozoic overburden. There, the reservoir (smV) and the barriers (so \& lcr) show no seismically detectable offsets. 

\subsection{Uncertainties of the fault interpretation}

The greatest uncertainties in the fault interpretation presented here are due to the low data density, sub-optimal orientation of the seismic profiles, and the general structural complexity. This difference in the seismic data, with sometimes massive differences in the visualisation of structures ultimately leads to an increased interpretational bias in the evaluation of the geometry and the orientation of individual faults. Fault patterns in the top of salt structures in particular can exhibit a complex interlocking of individual faults, as well with an overlapping of radial and circular fault patterns \citep{Jackson2017}. For this reason, the fault interpretation of closely spaced fault sets, as in the top of Henni, based solely on 2D seismic, leaves greater uncertainties in the linking of the individual fault picks from one seismic line to the next. This was taken into account in the interpretation by ensuring that the individual modeled faults show plausible displacements with the strike and that, without further indications, only fault picks with similar geometries and dips were linked to the depth without causing an excessive change in the fault's orientation. In a few cases, individual line interpretations had to be ignored to obtain a consistent fault geometry. In this context, the highest uncertainties in the current area are (1) at the southern flank and in the top of Henni South as well as (2) in the centre of the two structural cumulation points of Henni North and Henni South. In the latter, the faults in the top of Henni North interlock with those of Henni South. In this area, it is usually impossible to clearly assign individual 2D seismic fault picks to a fault plane. The SE- and ESE-striking faults on the southern flank of Henni are mainly only interpreted on the base of vintage data. Their different heterogeneous processing and generally lower data quality also lead to a greater interpretational bias for this area of the structure.

It should also be noted that the velocity model used is based on former modeling, in which the findings from the vintage data could not be taken into account. As a result, the velocity model used only provides a distorted representation of the velocity distribution with depth, particularly in the area of the southern flank of Henni. But the general error in the depth representation of horizons and faults is presumably significantly lower than the error caused by a general failure to take vintage data into account when modeling the Henni structure.

\subsection{Lithological description}

Based on the lithological descriptions provided in the drilling reports for salt pillow Henni South (H 15-1) and salt pillow Hella (H 9-1), descriptive and qualitative insights into the mechanical properties of the individual geological units can be derived.
In the Tertiary, Upper Miocene sands are generally unconsolidated, with a clayey matrix becoming more prominent toward the base. The Middle Miocene clays are described as soft and soft-plastic and variably silty, with localized firmer glauconitic material. The Lower to Middle Miocene features soft, plastic clays and sand. In the Lower Miocene, the sequence includes clays that are mostly soft, interspersed with thin, weakly consolidated sandy layers and occasional dolomitic stringers.

\begin{figure*}[t!]
\centering
{\includegraphics[trim=0 0 0 0, clip, width=0.78\linewidth]{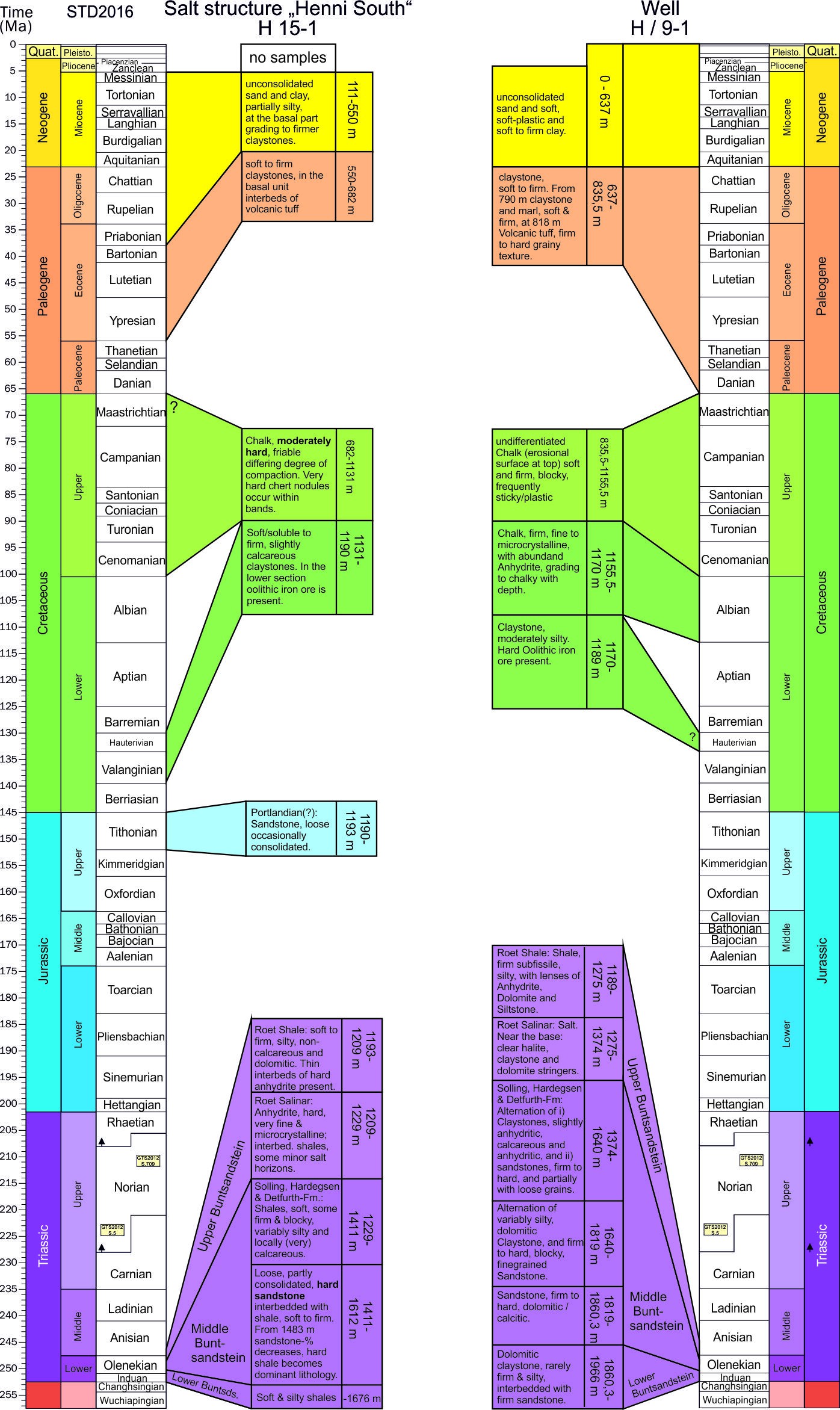}}
\caption{Qualitative descriptions of lithological \& mechanical properties in dependence of stratigraphy, based on the wells on Henni South (H 15-1) and salt pillow Hella (H 9-1).}
\label{fig:MechProps}
\end{figure*}

At Henni South, a gradual transition from unconsolidated clays to claystones is reported within a depth range of 505–550 m. It is associated with clays of the Eocene IV and therefore, first consolidated rocks are assigned to the sediments of Eocene III. In contrast, well H 9-1 exhibits a sharp boundary at a depth of 637 m, representing the base of the Lower Miocene. The Eocene rocks show a progression from very soft glauconitic clays in Eocene IV to slightly firmer claystones in Eocene II and III. Eocene I presents claystones that are soft to firm, with volcanic tuffs that are granular and blocky in texture. In the Cretaceous, the Upper Cretaceous chalks are moderately hard and friable, with nodular chert bands that are very hard and splintery. The Lower Cretaceous claystones are described as soft to firm, with the basal oolitic ironstone layer being more resistant. Jurassic sediments are only reported for Henni South, where (glauconitic) sandstones are occasionally consolidated but generally weakly cemented. In the Triassic, the Upper Buntsandstein is twofold and comprises soft to firm shales with interbedded hard anhydrite layers of the so-called Roet Shale and, of hard and microcrystalline anhydrite of the Roet salinar. The Middle Buntsandstein includes shales that are mostly soft but become blocky and firm in parts, with harder, well-cemented sandstones (fining-upward cycles). The Lower Buntsandstein or the Volpriehausen formation (lowest Middle Buntsandstein) is characterized by the alternation of shales and sandstones together with soft shales interspersed with thin, hard, calcareous oolitic sandstone layers. For Henni South at a depth of 1483 m, a decrease in sandstone percentage was observed and hard shale became the dominant lithology. At Hella (well H 9-1), the Volpriehausen (Fm) is characterized by the alternation of claystones, firm to hard sandstones, and interbedded oolites.

Based on the lithological descriptions from the drilling reports, the following purely descriptive and qualitative statements can be made regarding the lithological sequences:
\begin{enumerate}
    \item \textbf{Variability in consolidation}: The lithological units exhibit a wide range of consolidation states, from unconsolidated sands in the Upper Miocene to firm and hard layers such as the chalks of the Upper Cretaceous and the sandstones of the Triassic.
    \item \textbf{Soft and plastic clays}: Significant intervals, particularly in the Miocene and parts of the Eocene, are dominated by soft to plastic clays, often with minor silty or sandy components.
    \item \textbf{Chalk dominance in the Cretaceous}: The Upper Cretaceous sequence is characterized by moderately hard, friable chalks, occasionally interspersed with harder chert nodules.
    \item \textbf{Presence of firm claystones}: The Eocene contains increasingly firm claystones in the lower sections, transitioning from soft glauconitic clays.
    \item \textbf{Division in terms of lithology and mechanical properties}: Soft to firm shales with interbedded hard anhydrite layers representing the Roet shale, and hard \& microcrystalline anhydrite of the Roet salt.
    \item \textbf{Alternation of shales and hard sandstones Middle Buntsandstein}: The Triassic sequences of the Buntsandstein, include hard, well-cemented sandstones and distinct oolitic facies, contrasting with softer shales. Characteristic feature are fining upward cycles. 
    \item \textbf{Interbedding of soft and hard layers}: Many intervals, such as those in the Miocene, Eocene, and Triassic, show interbedding of soft clays and firmer or harder sandy or carbonate-rich layers, reflecting varied depositional environments.
    \item \textbf{Dolomitic and anhydritic layers}: Certain sections, particularly in the Triassic, include harder dolomitic and anhydritic interbeds, enhancing local lithological competence.
\end{enumerate}

\subsection{Stress state} \label{sec:stress state}

For this analysis, the orientation of the maximum horizontal stress for each fault section is derived through spatial interpolation from the 3D numerical stress model of Germany published by \cite{Ahlers2022}. The range of azimuth for the area is [340$^{\circ}$, 336$^{\circ}$] with a mean value of 338$^{\circ}$. We consider two scenarios of present-day stress magnitudes. In the first, the stress gradients are taken from the values measured in Kronprinzenkoog (\cite{Morawietz2020}) with maximum and minimum horizontal stress gradients of -19.90 MPa/km and -16.99 MPa/km, respectively, and a vertical stress gradient of -23.40 MPa/km, which are comparable with those reported by \cite{NOY2012} and \cite{FELLGETT2018} or close to those reported by \cite{ORLIC2016}. Additionally, the data from leak-off test collected from nearby exploration wells in the Dutch region (\url{www.nlog.nl}) show a mean gradient value of -16.6 MPa/km. The leak-off pressure, in many cases, can be used to estimate the magnitude of the minimum horizontal stress \citep{White2002}. These values are applied linearly from the surface to the depth under consideration, resulting in a normal faulting regime. In the second scenario, the stress tensors are taken directly from the numerical stress model by \cite{Ahlers2022}. The main difference here lies in the gradient of the maximum horizontal stress, which, in the second scenario, is higher than the gradient of the vertical stress for depths between zero to $\approx$2.5 km. The predominant regime thus is a strike-slip type since the fault data set considered in this study is no deeper than 3 km.

\section{Methods}\label{sec3}

The slip tendency for a cohesionless or non-virgin fault is defined according \cite{Morris1996} as
\begin{equation}
\label{eqn:slip1}
    T_S=\left| \frac{\tau}{\sigma'_n} \right|,
\end{equation}
which is based on Amonton's law or Coulomb-Mohr criterion on fault reactivation, which describes the initiation of a slip as
\begin{equation}
\label{eqn:MCcriterion}
    \tau \ge \tau_f=C_f- \mu_f~\sigma'_{n}.
\end{equation}
Here, $T_S$ is the slip tendency, $\tau$ and $\sigma'_n$ are the resolved shear stress and the effective normal stress acting along the fault's plane, respectively (\cite{Jaeger2011}), $\tau_f$ is the available frictional resistance of the interface, and $C_f$ and $\mu_f$ are the cohesion and the frictional coefficient, respectively. The effective stress tensor is defined as $\bm{\sigma}'=\bm{\sigma}+\bm{I}~P$, where $\bm{I}$ is the unit matrix and $P$ is the pore pressure. The normal and shear stresses for a given fault plane can be calculated from the acting principal stresses, see \cite{Jaeger2011}. We adopt in this contribution the engineering convention where compressive stresses and strains are negative. To improve readability, the slip tendency can be normalized with respect to the frictional coefficient, $\mu_f$, as (\cite{Peters2007}):
\begin{equation}
    \label{eqn:slip2}
    T_{S_{norm}}=\frac{T_S}{\mu_f}.
\end{equation}
Thus, a fault is in a critical condition when $T_{S_{norm}} \ge 1$.

The critical condition described above can also be presented in terms of a maximum sustainable pore pressure change. Such a presentation is important for planning and optimizing an injection strategy and potentially more appealing for decision makers or stakeholders in general. Eq. (\ref{eqn:MCcriterion}) can be written in terms of \textit{in-situ} pore pressure, $P_h$, and additional pore pressure, $\Delta_P$ (e.g., due to an injection) as
\begin{equation}
    \label{eqn:slip3}
    \tau \ge C_f - \mu_f~(\sigma_{n}+P_h+\Delta_P).
\end{equation}
A fault slip will be triggered when the pore pressure change reaches or exceeds the following:
\begin{equation}
    \label{eqn:slip4}
    \Delta_P \ge \frac{C_f-\tau}{\mu_f} - \sigma'_{n}.
\end{equation}
Thus, $\Delta_P$ can be defined as the maximum sustainable pore pressure change, which should not be exceeded to avoid bringing the fault to a critical state.
In eq. (\ref{eqn:slip4}), $\sigma'_n$ represents the initial effective normal stress, prior to an injection. Therefore, it should be noted that the above definitions do not consider the hydromechanical interaction taking place during the injection process. To consider such an interaction, the injection well's position, as well as the poroelastic properties of the region, must be known so that mechanical stress and pore pressure propagations can be correctly computed. In the absence of these data, the relation in eq. (\ref{eqn:slip4}) provides a rough estimation on the feasibility of an injection, from the perspective of a fault.

In this contribution, the slip-tendency analysis is carried out with the main purpose of deriving the likely focal mechanisms in the case that a microseismic event is triggered by the sequestration process. The friction coefficient, $\mu_f$, for each stratigraphic layer, is based on the study by \cite{SAMUELSON2012}. A value of 0.48 is assigned on faults located in layers dominated by fine-grained materials, i.e., in the Miocene\textendash Eocene, the Lower Cretaceous, the Upper Buntsandstein, the Lower Buntsandstein, and the upper layer of the Zechstein. The faults in the Upper Cretaceous are assigned $\mu_f$ of 0.6, based on the experimental result for Roet Claystone with higher quartz content \citep{SAMUELSON2012}. The upper part of the Middle Buntsandstein is dominated by fine-grained material, while its lower part has a relatively higher sandstone content (Fig. \ref{fig:MechProps}). For this layer, a value of 0.53 was obtained from averaging the values of all sub-layers proportional to their thickness in H 15-1.

Strikes and dips of the faults are calculated using their geometric vectors. The unit normal of each fault plane is enforced to have a positive vertical component so that the dip can be calculated correctly, and the strike is positioned accordingly. Rakes are then calculated based on the faults' striking direction and the shear stress tensor acting along the fault plane. The slip-tendency analysis for this study is written using MATLAB, while the fault model is developed using GOCAD.

To estimate the potential magnitudes, connected fault segments are clustered according to the stratigraphic strata and their orientation. In the latter, tolerance values of 5$^{\circ}$, 2$^{\circ}$, and 2$^{\circ}$ are applied for the strike, dip, and rake, respectively. The potential magnitude of each fault is then estimated using the scaling relation for fault area according to \cite{Kanamori2004}:
\begin{align}
    \label{eqn:magnitude}
    & M_w=\frac{{\log_{10}} M_0 -9.1}{1.5},
\end{align}
in which, in this case, the moment magnitude is linked to the pore pressure change in eq. (\ref{eqn:slip4}), such that
\begin{align} 
\label{eqn:magnitude2}
    & M_0=\Delta_P~A_f^{3/2}.
\end{align}
Here, $A_f$ is the collective area of each fault.

\begin{figure*}[t]
    \centering
        {\includegraphics[trim=0 0 0 0, clip, width=\linewidth]{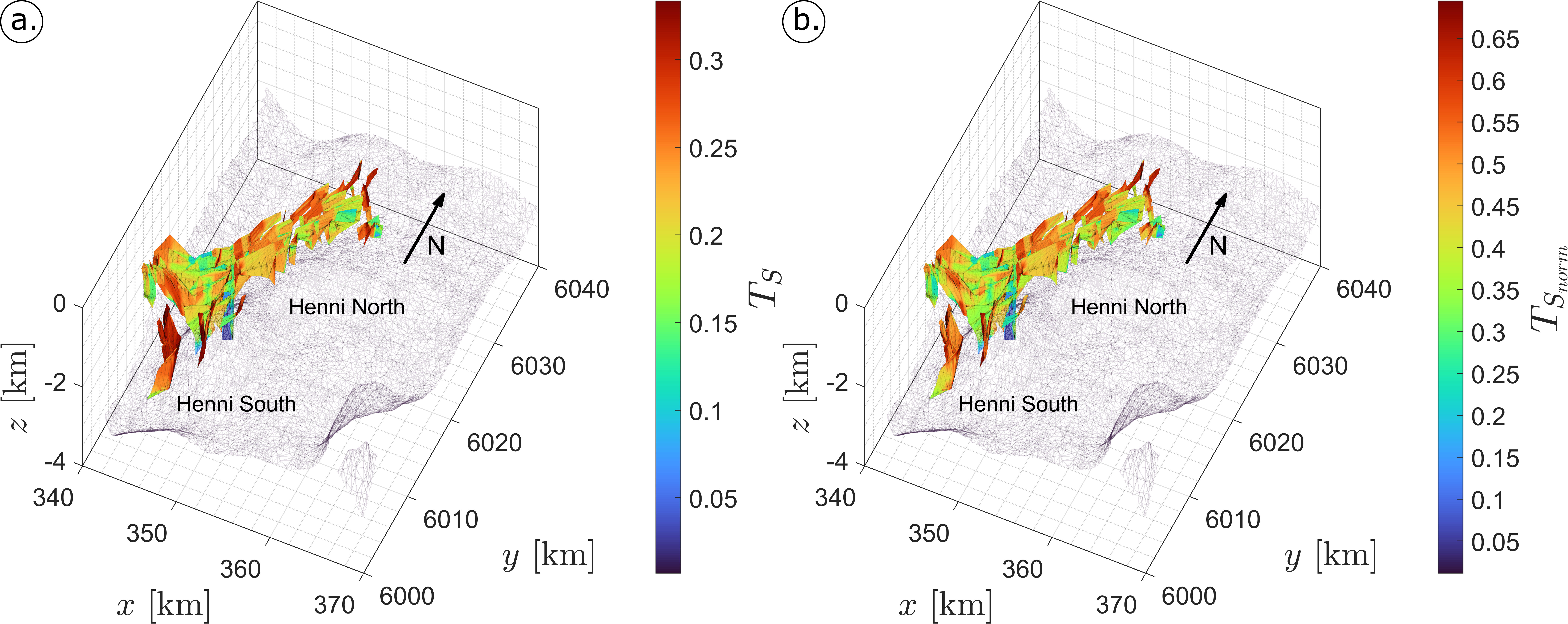}}
\caption{\textbf{(a)} Slip tendency, $T_{S_{}}$, and \textbf{(b)} its normalized value, $T_{S_{norm}}$, of the fault set near the targeted sub-traps assuming linear present-day stress gradients. In each sub-figure, the base of the Middle Buntsandstein is presented as a grey mesh.} 
\label{fig:Ts_Hor}
\end{figure*}

In this approach, it is assumed that the individual faults are non-interactive, including with their counterpart in the overburden or in the footwall in the case of a through-cutting fault nor with their parallel or crossing counterpart in the case of a multi-strand or interconnected fault. This assumption is valid when the permeability and mechanical properties permit. Another drawback is that eq. (\ref{eqn:magnitude2}) results in high potential moment magnitudes in faults with a high pore pressure window. Thus, faults with a high slip tendency tend to have a low potential moment magnitude. Furthermore, the calculation of the potential moment magnitude is based on the assumption that each fault in a given stratigraphic layer acts as a single continuity. In nature, slip along a rupture plane in the case of injection tends to start in an area where the stress change is highest (i.e., in the area of the fault that is directly in contact with a reservoir) and propagate to the nearby area of the fault's plane, see, e.g., \cite{Rutqvist2016}. To consider this aspect, an alternative scenario of eq. (\ref{eqn:magnitude2}), in which $A_f$ considers a height of $2\times h_r$, where $h_r$ is the height of the reservoir unit, is taken into account. According to H 15-1, $h_r$ is $\approx$40 m thick. A more detailed analysis, however, requires a proper definition of the poroelastic properties of each fault and is reserved for further study. 

\begin{figure*}[t! ]
    \centering
        {\includegraphics[trim=0 0 0 0, clip, width=\linewidth]{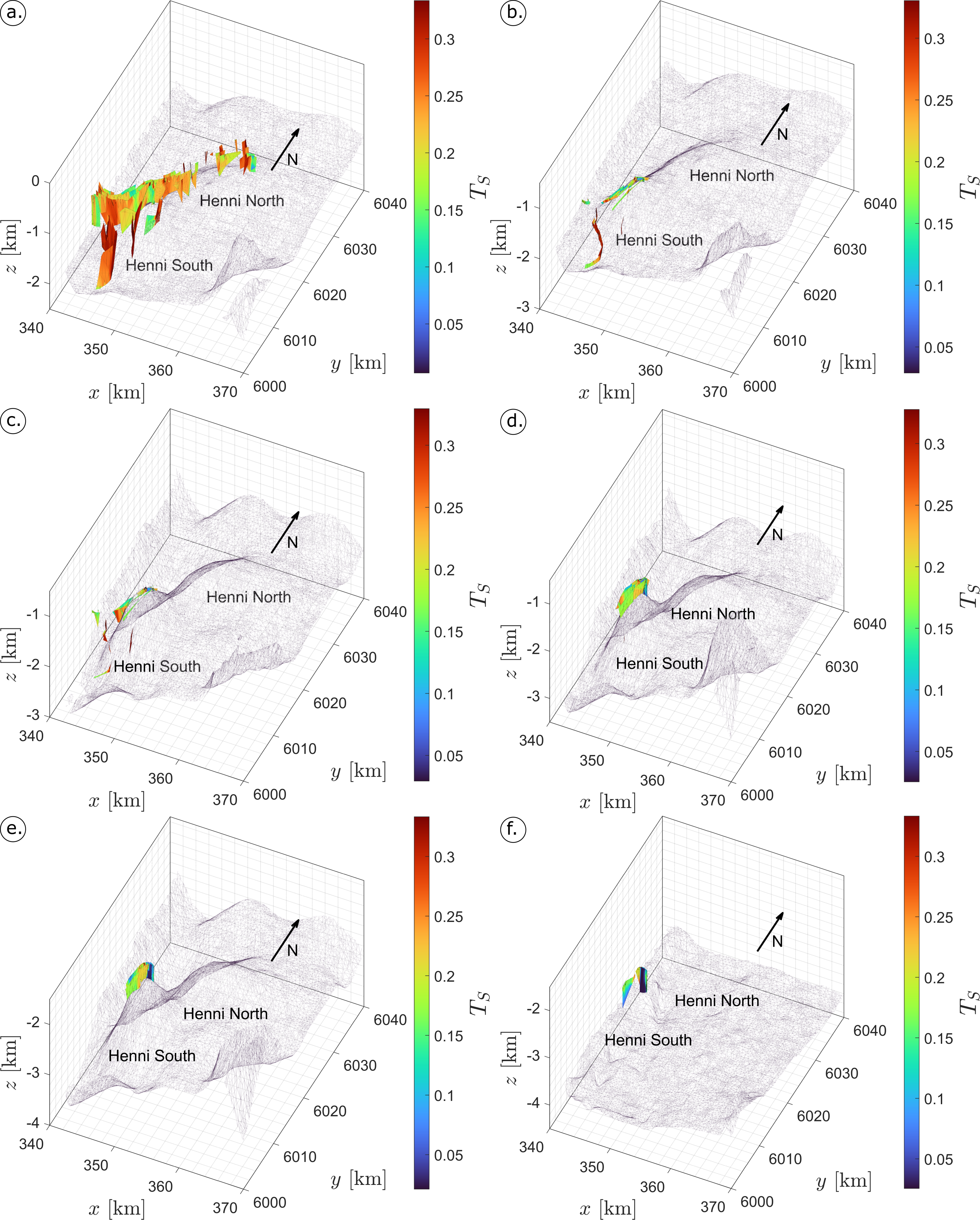}}
\caption{Slip tendency, $T_{S_{}}$, of the faults in \textbf{(a)} the Upper Cretaceous, \textbf{(b)} the Lower Cretaceous, \textbf{(c)} the Upper Buntsandstein, \textbf{(d)} the Middle Buntsandstein \textbf{(e)} the Lower Buntsandstein, and \textbf{(f)} the uppermost layer of the Zechstein (linear present-day stress gradients). In each sub-figure, the underlying base is shown as a grey mesh.} 
\label{fig:Ts_Hor_Cluster}
\end{figure*}

\section{Results and discussion}

\begin{figure*}[t!]
    \centering
        {\includegraphics[trim=0 0 0 0, clip, width=0.95\linewidth]{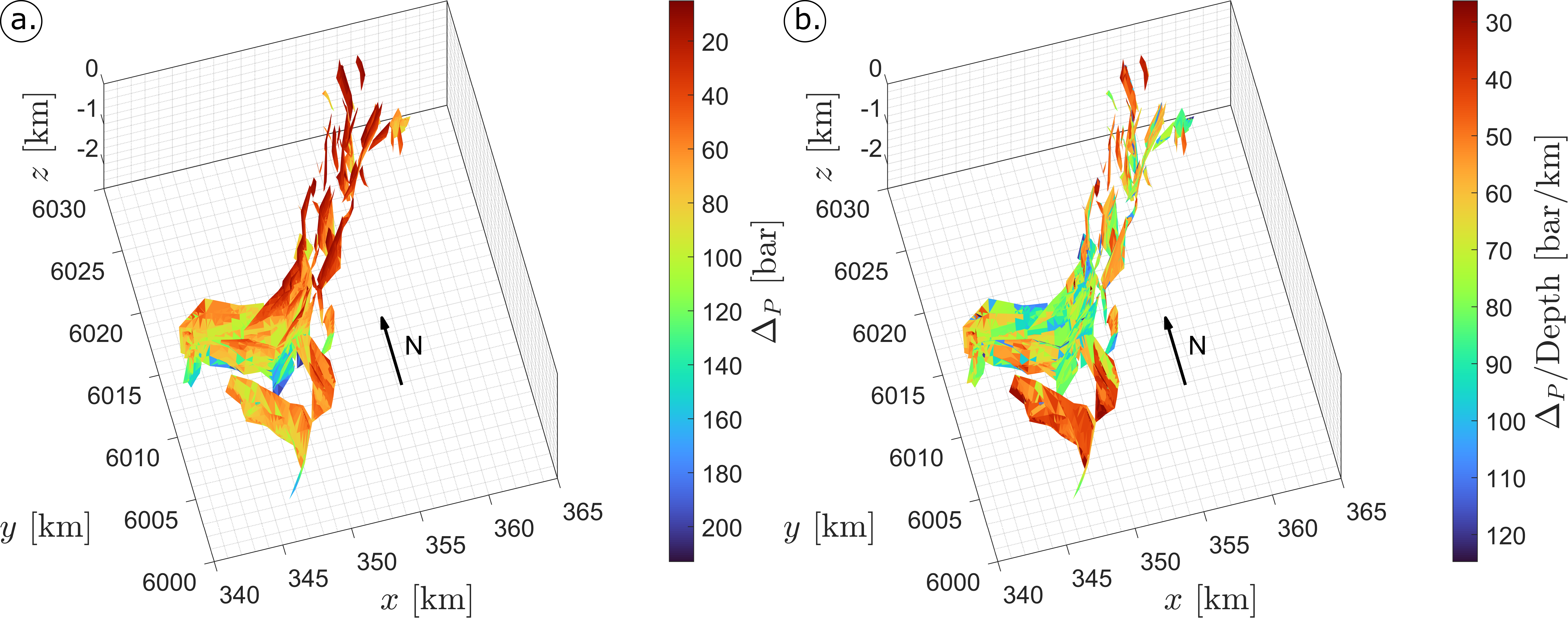}}
\caption{\textbf{(a)} Maximum sustainable pore pressure change, $\Delta_P$, and \textbf{(b)} its gradient, $\Delta_P$/Depth, of the fault set near the targeted sub-traps (linear present-day stress gradients).} 
\label{fig:DeltaP_DeltaPGrad}
\end{figure*}

The results of the slip-tendency analysis for the fault set near the targeted sub-traps assuming linear present-day stress gradients are presented in terms of $T_S$ and $T_{S_{norm}}$ in Fig. \ref{fig:Ts_Hor}. As both terms are linearly related, both sub-figures have a similar pattern and vary in scale. While the values of $T_S$ are in the range of [0.007, 0.334], those of $T_{S_{norm}}$ are in the range of [0.011, 0.695]. Most faults show a weak dependence of their slip tendency on the depth, while some faults exhibit such a strong dependence. In terms of location, critical faults with higher slip tendencies can be found at the northern and southern structures. The faults with NW/SE and NNW/SSE orientation show higher slip tendencies compared to those with other strikes, which corresponds to the azimuth of the maximum horizontal stress. Conversely, the lowest slip tendencies are observed along faults with NE/SW and ENE/WSW orientation, perpendicular to the aforementioned azimuth.

The slip tendencies of the faults, grouped according to stratigraphic strata, are presented in Fig. \ref{fig:Ts_Hor_Cluster}, which allows a more detailed overview of the values in each layer. Faults with $T_S>$0.31 can be found in the vicinity of Henni South within the Cenozoic, Mesozoic and Zechstein strata and above Henni North within the Cenozoic and Upper Cretaceous strata. The highest population of faults with $T_S>$0.31 is found in the Upper Cretaceous, followed by the Tertiary and the Lower Cretaceous. In the Upper Cretaceous (Fig. \ref{fig:Ts_Hor_Cluster}a), higher slip tendencies ($T_S>$0.32) are observed in the faults located along the axial region of the northern and southern Henni as well as at the southern and southwestern flank of Henni. These high values are found in the faults with strikes of [325$^{\circ}$, 355$^{\circ}$] and [140$^{\circ}$, 178$^{\circ}$]. Some faults in the west, south and southwest of Henni South extend to the Lower Cretaceous and the Upper Buntsandstein (Figs. \ref{fig:Ts_Hor_Cluster}a\textendash c, see also Section \ref{sec2}), and these transecting faults exhibit similar trend of slip tendency with a mean value of $\approx$0.2. Other transecting faults along the southern structure, extending until the Zechstein layer, exhibit mainly lower slip tendencies due to its orientation with mean values of 0.16, 0.15, and 0.15, respectively (Figs. \ref{fig:Ts_Hor_Cluster}d\textendash f).

\begin{figure*}[t!]
    \centering
        {\includegraphics[trim=0 0 0 0, clip, width=0.95\linewidth]{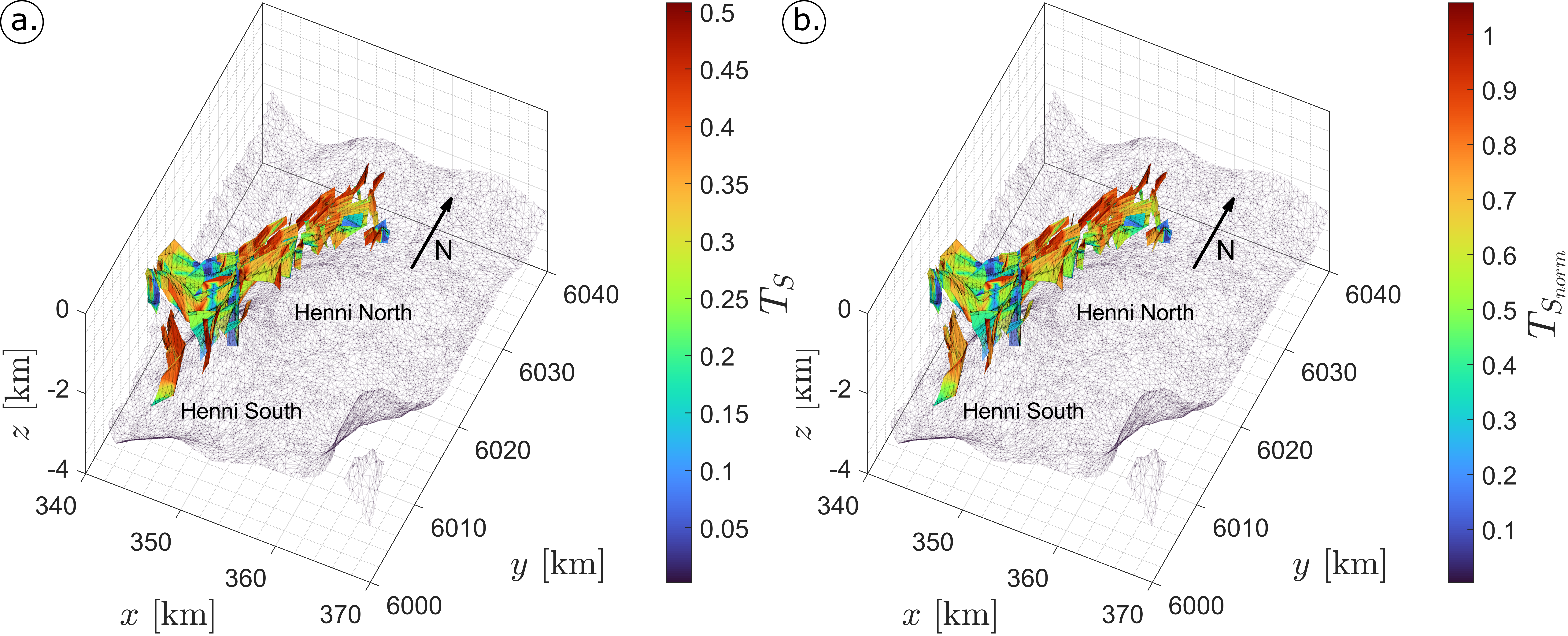}}
\caption{{\textbf{(a)} Slip tendency, $T_{S_{}}$, and \textbf{(b)} its normalized value, $T_{S_{norm}}$, of the fault set near the targeted sub-traps considering present-day stress tensors by \cite{Ahlers2022}. In each sub-figure, the base of the Middle Buntsandstein is presented as a grey mesh.}}
\label{fig:Ts_Hor_Ahlers}
\end{figure*}

\begin{figure*}[t!]
    \centering
        {\includegraphics[trim=0 0 0 0, clip, width=0.49\linewidth]{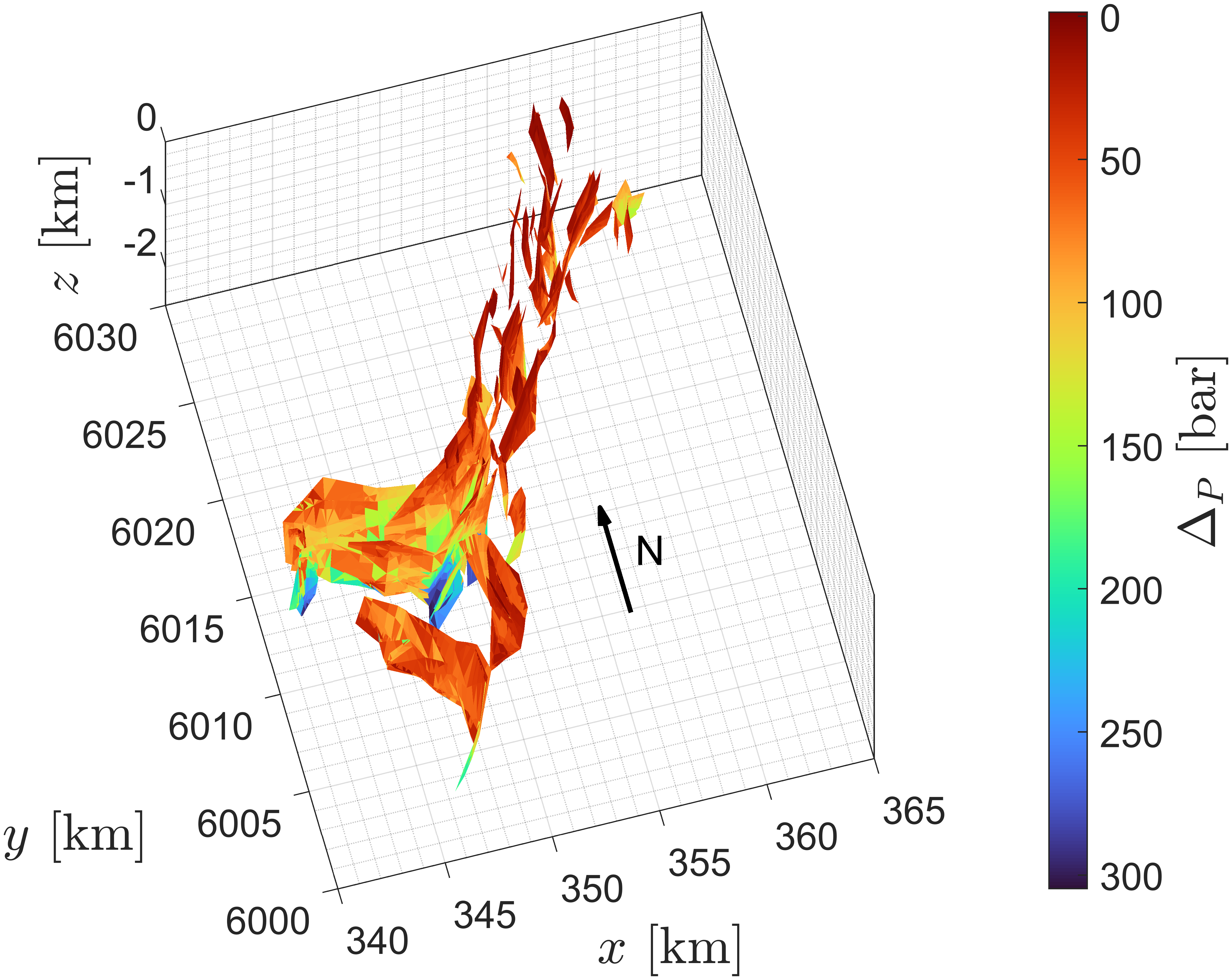}}
\caption{{Maximum sustainable pore pressure change, $\Delta_P$, of the fault set near the targeted sub-traps (present-day stress tensors according to \cite{Ahlers2022}).} }
\label{fig:DeltaP_Ahlers}
\end{figure*}

These results are then calculated as the maximum sustainable pore pressure change, $\Delta_P$, and its gradient, $\Delta_{P}$/Depth, which are shown in Fig. \ref{fig:DeltaP_DeltaPGrad}. The plot for $\Delta_P$ (Fig. \ref{fig:DeltaP_DeltaPGrad}a) shows a range of [4, 213] bar and a strong dependence of the value on the depth. The critical values are observed on the near-surface faults, while high pressure windows are observed in the deeper faults, which is due to the higher confining stresses. When computed as gradients over the depth, the values are in the range of [26, 125] bar/km (Fig. \ref{fig:DeltaP_DeltaPGrad}b), and the distribution pattern changes significantly. The pattern is now closer to the one of the slip-tendency, where critical values are observed in the faults whose strikes are close to the azimuth of the maximum horizontal stress. Under the considered assumptions, the minimum tolerable pore pressure change of the faults in the Middle Buntsandstein is $\approx$45 bar. A continuation of this study, in which the pore pressure changes caused by various CO\textsubscript{2} injection schemes are evaluated with regard to the results presented here, is ongoing.

{The slip tendencies and its normalized values considering present-day stress tensors by \cite{Ahlers2022} are shown in Fig. \ref{fig:Ts_Hor_Ahlers}. The maximum $T_S$ in this case is 0.51 while the maximum $T_{S_{norm}}$ is higher than 1. When $\mu_f$ of the fault set is increased to 0.57 following the approach by \cite{Roeckel2022}, the maximum $T_{S_{norm}}$ is $\approx$0.9. The distribution pattern of the slip tendencies are quite close to the one obtained using linear present-day stress gradients (Fig. \ref{fig:Ts_Hor}). Although the higher $T_{S_{norm}}$ values here indicate that the faults are critically oriented and stressed, especially in the younger strata, they do not necessarily translate into seismic activity, as this depends on the dominant material type in the fault zone and the rheological property of the frictional strength of the fault, e.g., see \cite{Scholz1998, SAMUELSON2012}. The $\Delta_{P}$ computed using this assumption are shown in Fig. \ref{fig:DeltaP_Ahlers}. The stark difference in the range of $\Delta_{P}$, compared to those in Fig. \ref{fig:DeltaP_DeltaPGrad}a, is caused by the changes in the normal and shear stresses acting on the faults, which are also reflected in the changes of the rakes. For example, the fault with a $\Delta_{P}$ of 300 bar in Fig. \ref{fig:DeltaP_Ahlers} has a rake of 42$^{\circ}$, indicating a strike-slip or near reverse movement tendency, while the same fault in Fig. \ref{fig:DeltaP_DeltaPGrad}a exhibits a $\Delta_P$ of 166 bar and has a rake of -88$^{\circ}$. When using this assumption, the minimum $\Delta_{P}$ of the faults within the Middle Buntsandstein decreases to $\approx$17 bar, indicating a significantly lower sustainable injection rate or a requirement to rearrange the injection scheme altogether.}

\subsection{Influence of cohesion}

\begin{figure}
    \centering
        {\includegraphics[trim=0 0 0 0, clip, width=0.6\linewidth]{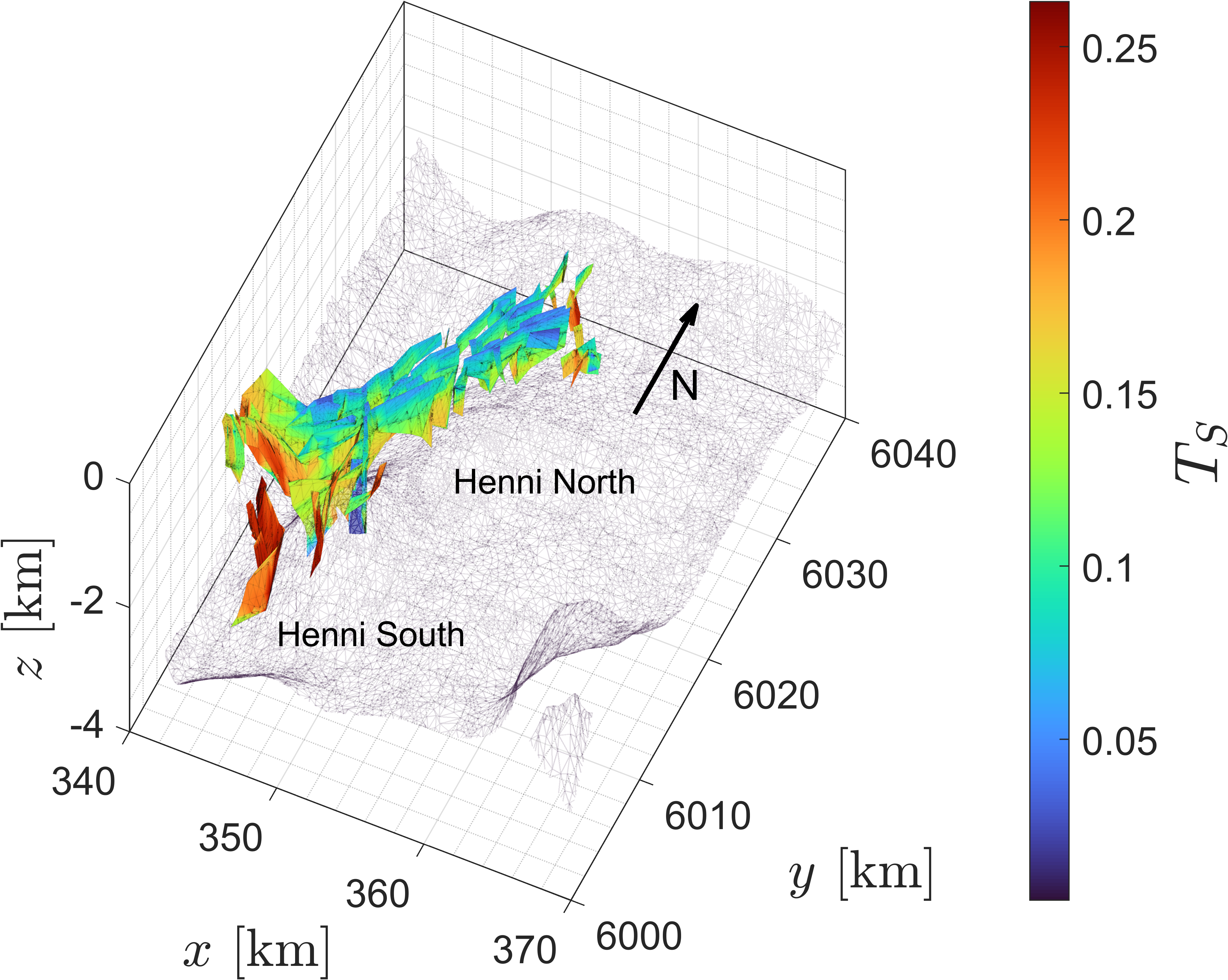}}
\caption{Slip tendency, $T_{S_{}}$, of the fault set near the targeted sub-traps considering a cohesion of 2.5 MPa (linear present-day stress gradients).} 
\label{fig:TsNorm_coh2p5}
\end{figure}

\begin{figure*}[t!]
    \centering
        {\includegraphics[trim=0 0 0 0, clip, width=0.95\linewidth]{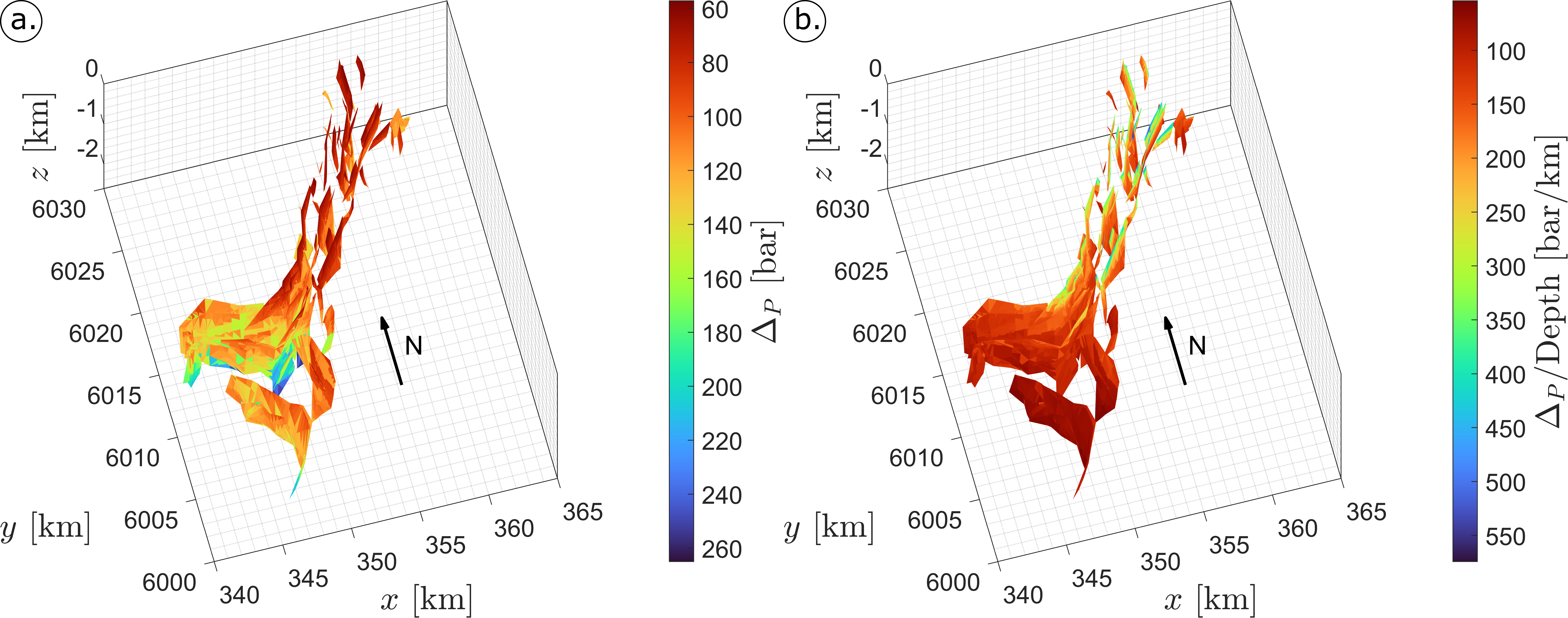}}
\caption{\textbf{(a)} Maximum sustainable pore pressure change, $\Delta_P$, and \textbf{(b)} its gradient, $\Delta_P$/Depth, of the fault set near the targeted sub-traps considering a cohesion of 2.5 MPa (linear present-day stress gradients).} 
\label{fig:DeltaP_DeltaPGrad_coh2p5}
\end{figure*}

To investigate the influence of cohesion on the slip tendency, a calculation considering a cohesion of 2.5 MPa is performed and the results are shown in terms of $T_{S_{}}$ in Fig. \ref{fig:TsNorm_coh2p5} and $\Delta_P$ and $\Delta_P$/Depth in Fig. \ref{fig:DeltaP_DeltaPGrad_coh2p5}. A cohesion value of 2.5 MPa is relatively low given that the region is seismically quiet. It is likely that cementation has taken place along the fault within the Pre-Cretaceous strata and cohesion along these faults' plane is higher than this value. As an example, a study by \cite{Reyer2014} showed high unconfined compressive strengths in samples of Triassic sandstone, indicating high cohesion. However, \cite{Jaeger2011} suggested to consider no or strongly reduced cohesion for rocks in the vicinity of fractures. Thus, considering that primary data is not yet available, this value is taken to qualitatively assess the influence of the parameter.

Fig. \ref{fig:TsNorm_coh2p5} shows that the improvement due to cohesion is stronger for the shallow-depth faults, which is due to the constant value of cohesion assumed for this case study. Compared to those in Fig. \ref{fig:Ts_Hor}, the results in this case show a higher dependence on the depth. The plot for $\Delta_P$ (Fig. \ref{fig:DeltaP_DeltaPGrad_coh2p5}a) shows the same pattern as that found in the comparable sub-figure in Fig. \ref{fig:DeltaP_DeltaPGrad}. The range of the values is linearly increased by 2.5/$\mu_f$ MPa to [56, 265] bar. The values of $\Delta_P$/Depth are in the range of [53, 573] bar/km (Fig. \ref{fig:DeltaP_DeltaPGrad_coh2p5}b), and the corresponding plot shows, in addition to the dependence on the fault geometry, a stronger dependence of the variable on the depth compared to the results of the calculation with no cohesion. Particularly for the Middle Buntsandstein, the consideration of a cohesion of 2.5 MPa increased significantly the $\Delta_P$ to $\approx$92 bar. 

\subsection{Influence of strike, dip, and rake}

\begin{figure*}
    \centering

        {\includegraphics[trim=0 0 0 0, clip, width=\linewidth]{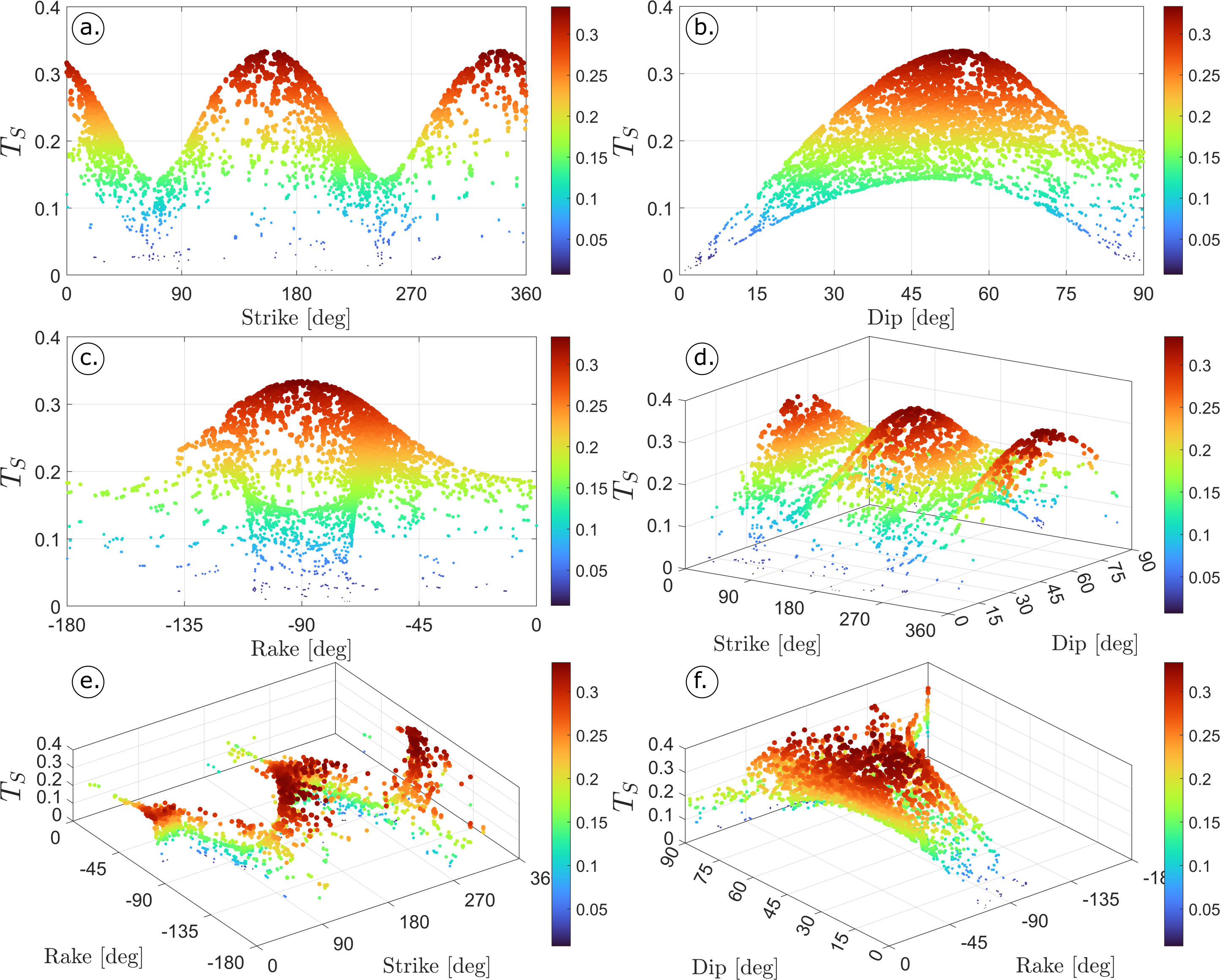}}

\caption{Slip tendency, $T_{S_{}}$, of the fault set near the targeted sub-traps in relation to its \textbf{(a)} strike, \textbf{(b)} dip, \textbf{(c)} rake, and in cross relation to \textbf{(d)} strike and dip, \textbf{(e)} strike and rake, and \textbf{(f)} dip and rake (linear present-day stress gradients).}
\label{fig:TsRel}
\end{figure*}

\begin{figure*}
    \centering
        {\includegraphics[trim=0 0 0 0, clip, width=0.95\linewidth]{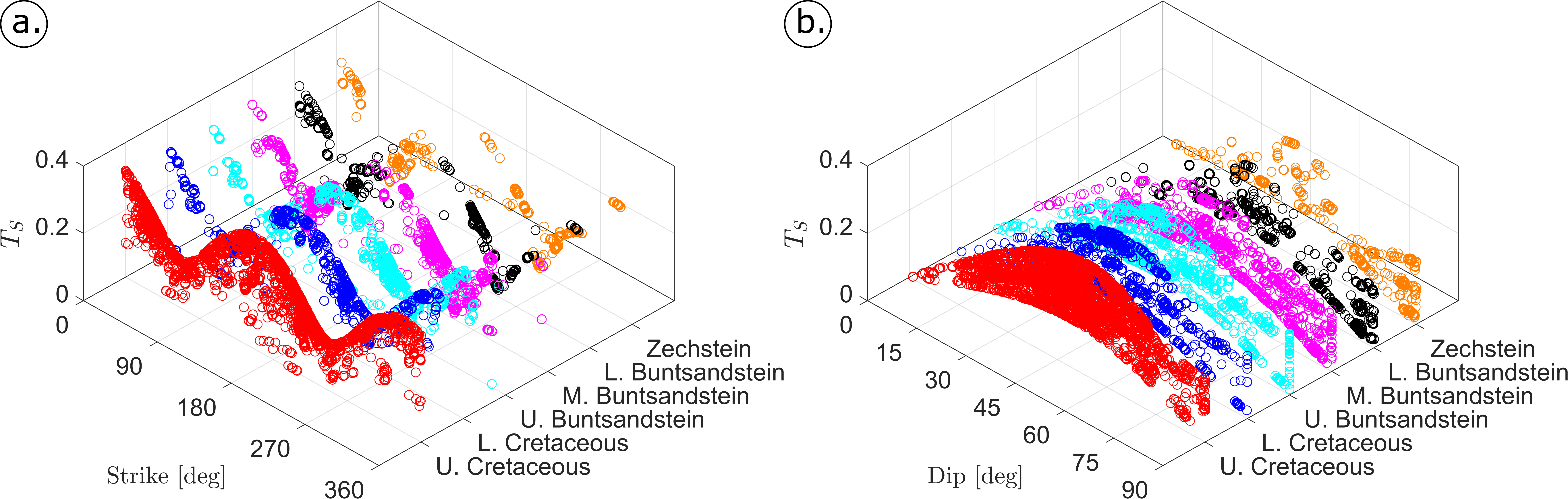}}
\caption{Slip tendency, $T_{S_{}}$, of the fault set near the targeted sub-traps, clustered according to the stratigraphic units, in relation to the faults' \textbf{(a)} strike and \textbf{(b)} dip (linear present-day stress gradients).} 
\label{fig:Ts_MultiLayer}
\end{figure*}

To show the vulnerable focal mechanisms under the assumption of linear present-day stress gradients, the slip tendencies are plotted in relation to their strike, dip, and rake in Fig. \ref{fig:TsRel}. The highest $T_S$ of 0.334 is observed in a fault segment with a strike of 340$^{\circ}$ (Figs. \ref{fig:TsRel}a), while $T_S$ of 0.333 is also observed in a fault with a strike of 155$^{\circ}$. A deviation of 20$^{\circ}$ from these critical strikes results in a reduction of the $T_S$ by $\approx$5\%. The critical dip can be identified in Figs. \ref{fig:TsRel}b as 55$^{\circ}$, and a $\approx$5\% reductions are observed in -10$^{\circ}$ and +7$^{\circ}$ deviation from the critical point. In terms of rake, Fig. \ref{fig:TsRel}c shows that the critical value is -90$^{\circ}$ and a deviation of 16$^{\circ}$ from this value results in a reduction of $T_S$ by $\approx$5\%. These critical strike, dip, and rake occur at the same or similar fault segment, as shown in cross plots relating $T_S$ to the strikes, dips, and rakes in Figs. \ref{fig:TsRel}d-f. The cross plot also shows that the slip tendencies of faults with dips between [15$^{\circ}$, 65$^{\circ}$] follow a double-saddle shape in relation to the strike parameter. In general, the patterns observed in the results are in line with the common theory of slip mechanism, e.g., see \cite{Scholz_2019}. In addition, the identified critical focal mechanisms relate quite well with two seismic events cataloged by \cite{SHARP2022} for the Central North Sea, whose strikes, dips, and rakes are \textit{(1)} 356$^{\circ}$, 85$^{\circ}$, -95$^{\circ}$; and \textit{(2)} 142$^{\circ}$, 35$^{\circ}$, -80$^{\circ}$. Besides the apparent similarity, it should be noted that these cataloged seismic events are located in an area whose tectonic setting is not entirely comparable to the one of Henni.

The slip tendencies, in relation to the faults' strike and dip, are plotted according to the stratigraphic units in Fig. \ref{fig:Ts_MultiLayer}. The figure shows that the faults in the Upper Cretaceous, Lower Cretaceous, and the Upper Buntsandstein have a wide range of strikes and dips (Fig. \ref{fig:Ts_MultiLayer}a). In these layers, the faults along the northern and southern sub-traps show a mean strike value of 53$^{\circ}$ (after reversal of the conjugate faults), in line with the direction of the two maxima along the salt pillow (see Fig. \ref{fig:HenniFaultTopview}), and a mean $T_S$ of 0.22. In the southwestern flank of Henni South, the faults are oriented towards a mean strike of 104$^{\circ}$, which results in a mean $T_S$ of 0.23. The strike characterization of the faults in the three units of Buntsandstein shows a gradual tendency over the depth towards NE with mean values of 75$^{\circ}$, 66$^{\circ}$, and 65$^{\circ}$ (after reversal) for the upper, middle and lower parts, respectively. This leads to relatively low $T_S$ values with a mean of 0.13\textendash0.17 in all three layers. However, several fault segments are oriented towards the critical strike and dip and, therefore, exhibit above-average slip tendencies. Similarly, the dip characterization of the fault set in the Upper Cretaceous, Lower Cretaceous, and the Upper Buntsandstein as shown in Fig. \ref{fig:Ts_MultiLayer}b shows a relatively consistent distribution with mean values of 49$^{\circ}$\textendash53$^{\circ}$. In the subsequent Buntsandstein layers and below, the number of low-dipping faults gradually decreases over the depth, as shown by the increasing mean values of 61$^{\circ}$, 67$^{\circ}$, and 68$^{\circ}$, respectively (see also Fig. \ref{fig:CrossSeismic}).

\begin{figure*}
    \centering

        {\includegraphics[trim=0 0 0 0, clip, width=\linewidth]{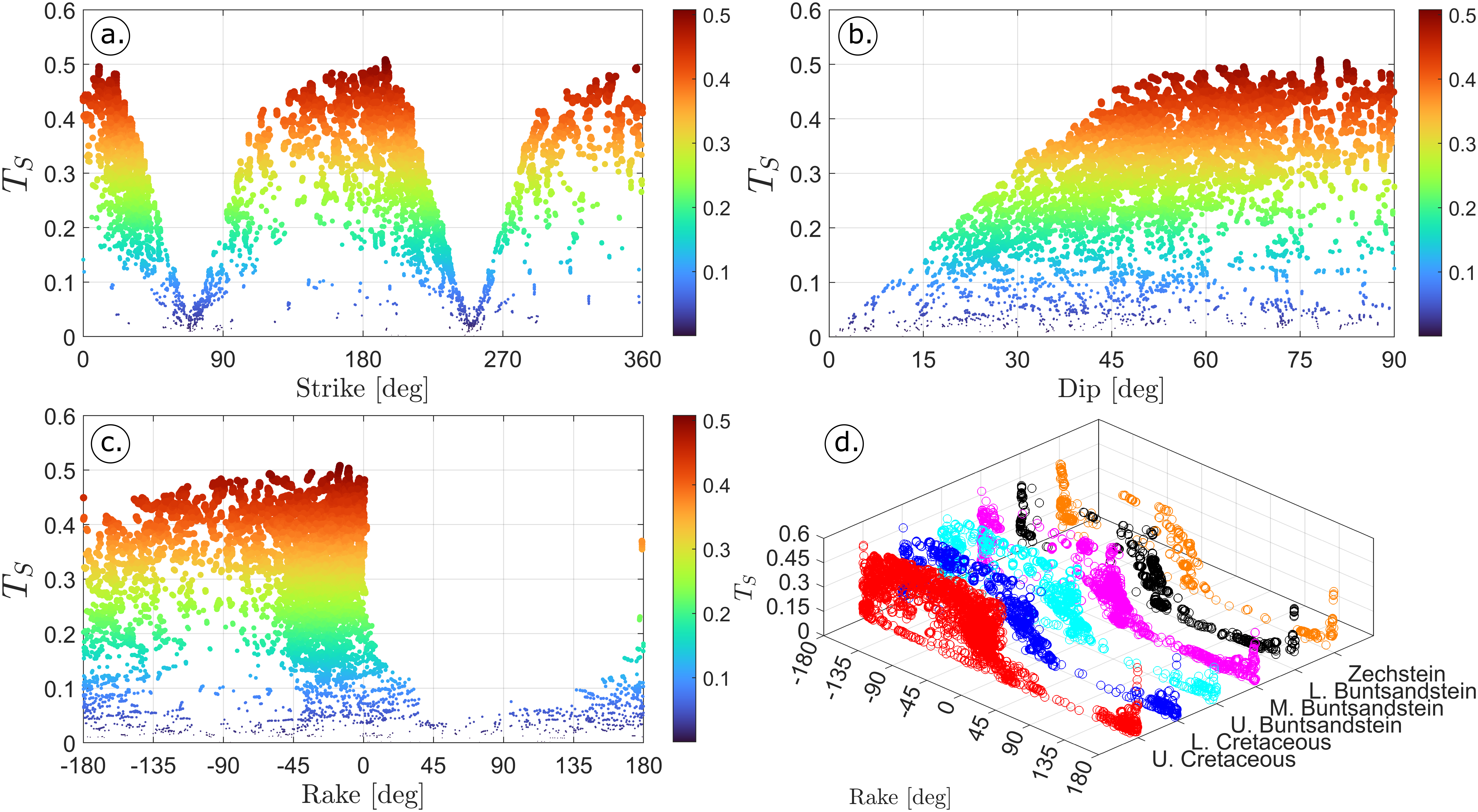}}

\caption{{Slip tendency, $T_{S_{}}$, of the fault set near the targeted sub-traps in relation to its \textbf{(a)} strike, \textbf{(b)} dip, \textbf{(c)} rake, and \textbf{(d)} 
clustered according to the stratigraphic units and in relation to rake (present-day stress tensors according to \cite{Ahlers2022}).}}
\label{fig:TsRel_Ahlers}
\end{figure*}

{The relations between the slip tendencies with the fault geometry and movement tendency obtained from the analysis considering stress tensors according to \cite{Ahlers2022} are presented in Fig. \ref{fig:TsRel_Ahlers}. In this case, slip tendencies $>$0.48 can be observed in faults whose strikes are in the range of [156$^{\circ}$, 197$^{\circ}$] and [356$^{\circ}$, 23$^{\circ}$] (Fig. \ref{fig:TsRel_Ahlers}a). The ranges of critical strikes are significantly wider than the ones obtained using linear stress gradients in Fig. \ref{fig:TsRel}a. Similarly, the critical dips and rakes have wider ranges with slip tendencies $>$0.48 shown in the faults with dips of [53$^{\circ}$, 90$^{\circ}$] (Fig. \ref{fig:TsRel_Ahlers}b) and rakes of [-85$^{\circ}$, 0$^{\circ}$] (Fig. \ref{fig:TsRel_Ahlers}c). Since this study considers faults that are no deeper than 3 km, the stress tensors by \cite{Ahlers2022} result in a dominant strike-slip type of fault, as discussed in Section \ref{sec:stress state} and shown in Fig. \ref{fig:TsRel_Ahlers}d. The critical focal mechanisms, especially the rakes, and the predominant fault type do not match very well with the two cataloged events mentioned before \citep{SHARP2022} and the fault offsets observed in Fig. \ref{fig:CrossSeismic}, which mainly show normal faults, even at the base of MMU. However, the latter is likely due to a modification in the local stress regime due to the local geological setting and history. The deformation and faulting in the area are more likely driven by the upwelling of the Henni salt structure rather than the far-field stresses. Therefore, a further field investigation is required to confirm the local current stress magnitudes and orientation as well as the frictional and poroelastic properties of the faults in order to draw a more solid conclusion. }

\subsection{Potential magnitude of seismic events}

\begin{figure*}[t!]
    \centering
        {\includegraphics[trim=0 0 0 0, clip, width=\linewidth]{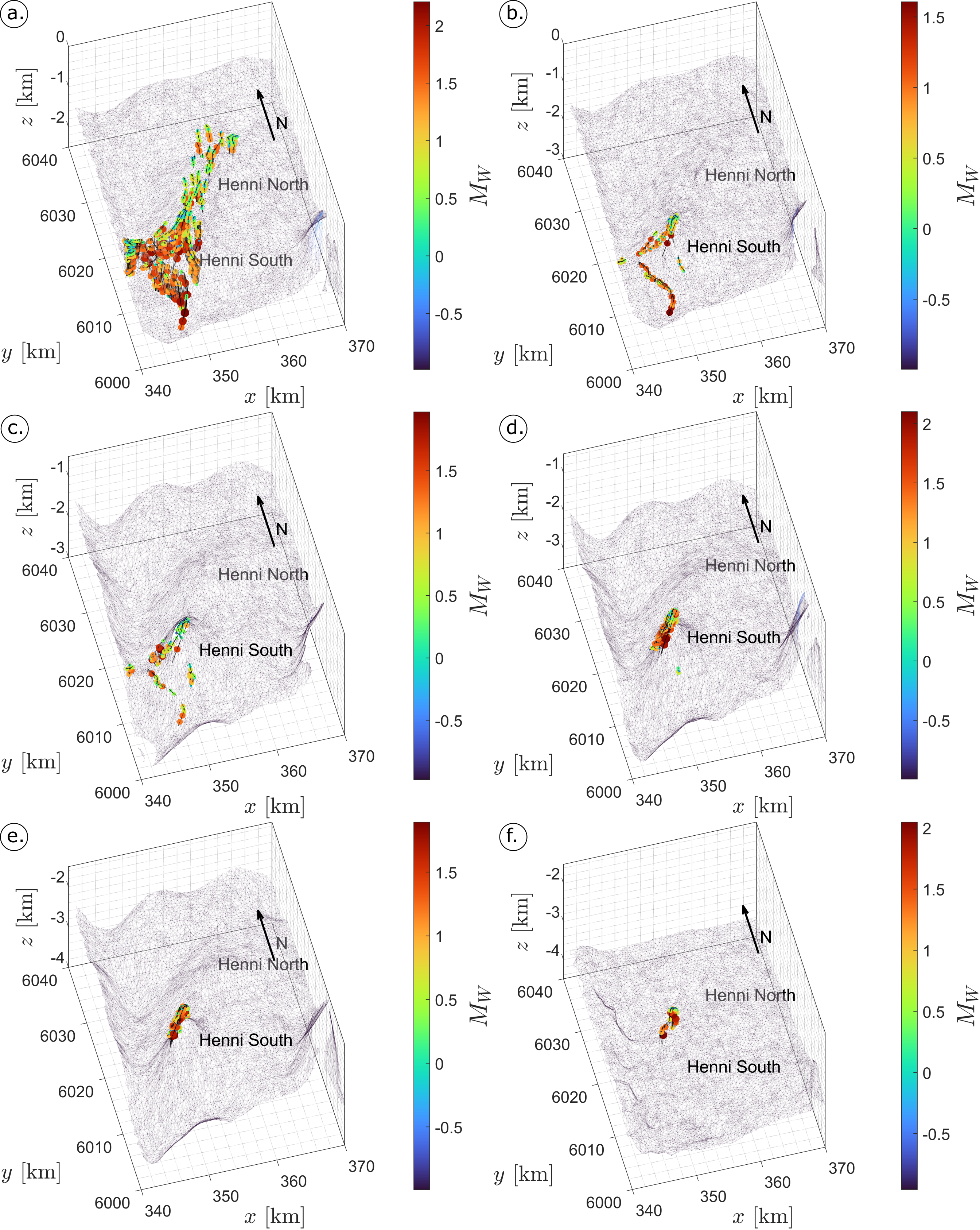}}
\caption{Potential moment magnitudes ($M_W>-1$) of the faults near the targeted sub-traps in \textbf{(a)} the Upper Cretaceous, \textbf{(b)} the Lower Cretaceous, \textbf{(c)} the Upper Buntsandstein, \textbf{(d)} the Middle Buntsandstein, \textbf{(e)} the Lower Buntsandstein, and \textbf{(f)} the uppermost layer of the Zechstein (linear present-day stress gradients). In each sub-figure, the corresponding base is shown as a grey mesh.}
\label{fig:Mw_Combined}
\end{figure*}

\begin{figure}
    \centering
        {\includegraphics[trim=0 0 0 0, clip, width=0.9\linewidth]{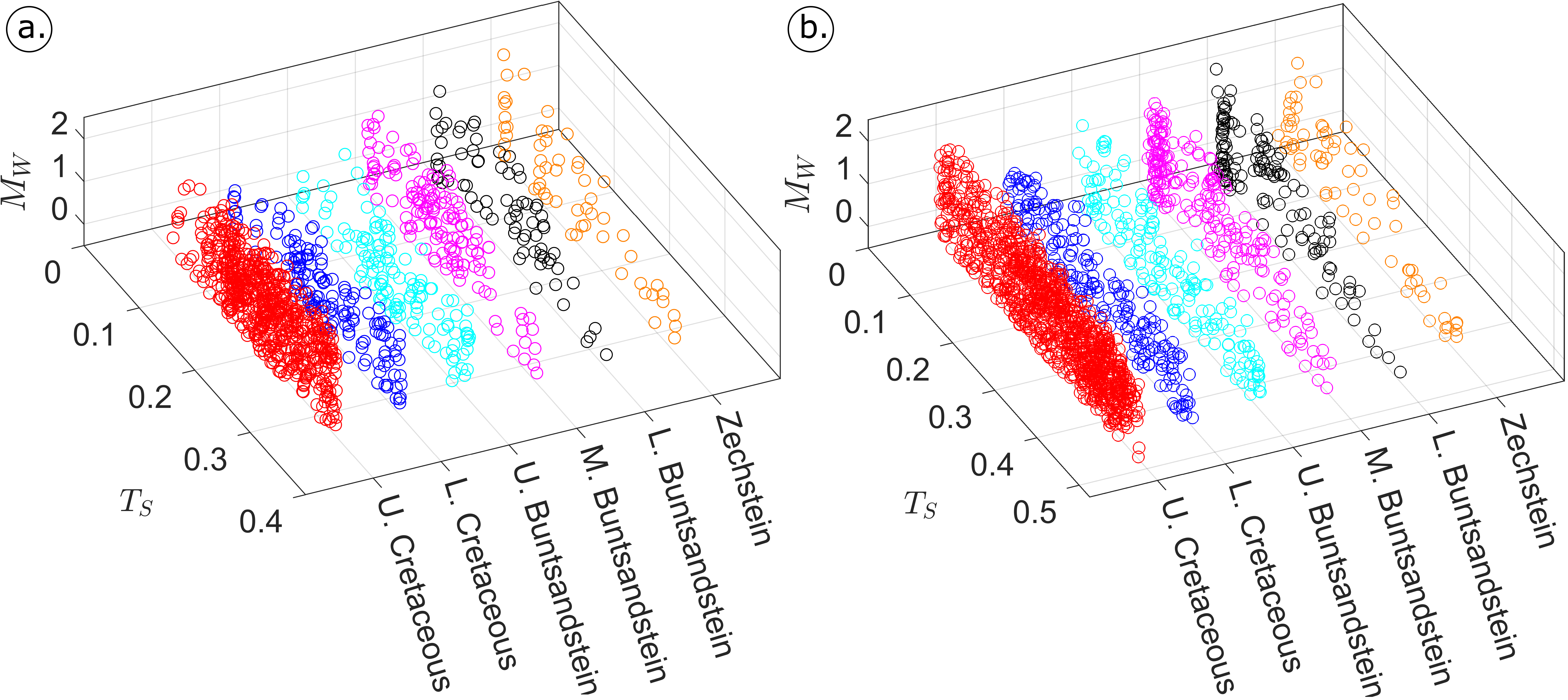}}
\caption{Potential moment magnitudes, $M_{W_{}}$, of the fault set near the targeted sub-traps, clustered according to the stratigraphic units and the orientation and potential movement of the faults, plotted in relation to its slip tendencies, $T_S$, obtained \textbf{(a)} using the assumption of linear present-day stress gradients and \textbf{(b)} considering the stress tensors by \cite{Ahlers2022}.} 
\label{fig:Mw_Ts_MultiLayer}
\end{figure}

The potential magnitudes of seismic events of the fault set, calculated using eq. (\ref{eqn:magnitude2}) assuming linear present-day stress gradients, are shown for different stratigraphic layers in Figs. \ref{fig:Mw_Combined} and \ref{fig:Mw_Ts_MultiLayer}. Only faults with potential magnitude $>$-1 are shown for clarity. For this calculation, a cohesion of zero is assumed and the potential moment magnitude of each fault is calculated assuming a continuous rupture plane along each stratigraphic layer (see Section \ref{sec3}). 

In the Upper Cretaceous, several faults in the vicinity of the southern sub-trap show a combination of elevated slip tendencies and elevated potential magnitudes (Fig. \ref{fig:Mw_Combined}a). For example, at the southern flank of Henni South, several faults show $T_S$ of [0.25, 0.31] and potential moment magnitudes of [1.83, 2.20] $M_W$. In these cases, the combination of above-average values of slip tendency and potential moment magnitude is the direct result of the faults' orientation and respective area. These faults are separated from the reservoir unit by the main seal. However, a potential reactivation must not be ruled out without further information.

At Henni North, 9 faults with potential magnitudes of 1.42$>M_W>$1.65 are observed in the Upper Cretaceous (Fig. \ref{fig:Mw_Combined}a). Some of these faults are oriented at strikes of 351$^{\circ}$ and 2$^{\circ}$, which are relatively close to the azimuth of the maximum horizontal stress, while the rest are oriented at strikes of 32$^{\circ}$\textendash 60$^{\circ}$. The most unfavorable combination of these faults is the one characterized by a strike of 2$^{\circ}$ and a dip of 52$^{\circ}$, which results in a $T_S$ of 0.31 and a potential event of 1.42 $M_W$.

In the Middle Buntsandstein, in which the reservoir unit is located, two faults with potential moment magnitudes $>$2.0 $M_W$ at Henni south are observed (Fig. \ref{fig:Mw_Combined}d). Within this layer, a direct migration of stresses, strains, and pore pressures is expected in the case of an injection. These faults have strikes and dips of (215 and 230)$^{\circ}$ and (31 and 47)$^{\circ}$, respectively, and a slip tendency of $\approx$0.17. Thus, these faults are not critically oriented and stressed. On the other hand, slip tendencies $>$0.31 are observed on faults with strikes of 150$^{\circ}$\textendash175$^{\circ}$ and dips of 45$^{\circ}$\textendash58$^{\circ}$, and they show potential moment magnitudes of (-0.18)\textendash0.62 $M_W$.

The fault segments of the same fault clusters at Henni South, crossing into the Upper Buntsandstein and the Lower Cretaceous have maximum potential moment magnitudes of 1.97 and 1.61, respectively (Fig. \ref{fig:Mw_Combined}d and c), while the ones in the Lower Buntsandstein and the upper layer of the Zechstein have potential magnitudes of up to 1.89 and 2.05, respectively (Figs. \ref{fig:Mw_Combined}e and f). As observed in the Middle Buntsandstein, the faults related to these values in these layers are not critically oriented and stressed. Among the faults with $T_S>$0.30, the maximum potential moment magnitudes are 1.28 $M_W$, 1.48 $M_W$, 0.81 $M_W$, 0.07 $M_W$, and 0.54 $M_W$ in the Lower Cretaceous, the Upper Buntsandstein, the Middle Buntsandstein, the Lower Buntsandstein, and the upper layer of the Zechstein, respectively. Further examination into the Cenozoic overburden reveals a potential moment magnitude of 0.93 $M_W$ in the critically oriented faults. Thus, the combination of above-average potential moment magnitudes and slip tendencies is only found in the upper part of Mesozoic near Henni South, which can be related to the larger offsets found in this layer (see Section \ref{sec2 inventory}). The clear trend of decreasing $M_W$ as $T_S$ increases in the Buntsandstein is further illustrated in Fig. \ref{fig:Mw_Ts_MultiLayer}a, which also shows the more complex relation between the variables in the upper part of the Mesozoic. The presence of larger faults with critical orientation in these layers results in above-average slip tendencies and potential magnitudes. Similar trends are also observed when the fault set is analyzed considering the stress tensors by \cite{Ahlers2022} (Fig. \ref{fig:Mw_Ts_MultiLayer}b).

Due to the assumption of a continuous rupture plane at each fault, the above estimated moment magnitudes tend to be on the conservative side. Using an alternative assumption that the height of the rupture plane is limited to $2 \times h_r$, where $h_r$ is $\approx$40 m according to well H 15-1, the potential moment magnitudes of the faults in the Middle Buntsandstein at Henni South are [-5.39, 1.39] $M_W$. The fault with the highest potential moment magnitude is not critically oriented and stressed ($T_S$=0.18). Among the faults with $T_S>$0.25, the potential moment magnitudes are [-2.27, -0.07] $M_W$. However, in this case, the minimum $\Delta_P$ is slightly reduced to $\approx$42 bar.

\section{Conclusions}

Following a previous study which shows potential saline aquifer storage sites on the West Schleswig Block within the German North Sea \citep{FUHRMANN2024}, we investigate the stability of existing faults in a selected study area to reactivation. The fault interpretation and modeling are based on 80 2D seismic data combining 60 years of exploration history. The geological history, features, and lithological description of the area are discussed. A slip-tendency analysis is then performed using the 3D fault model and the 3D stress model of Germany, and it reveals the critical focal mechanisms, which are well correlated with back-calculated events in the Central North Sea \citep{SHARP2022}. The slip-tendency procedure is also extended to derive the maximum sustainable pore pressure change which is important in defining injection strategies. Furthermore, the potential moment magnitude in the event of a fault reactivation is estimated using an assumption that the change of stress would not exceed the triggering pore pressure change.

When assuming linear present-day stress gradients with a normal-faulting regime, the results of the slip-tendency analysis show that the faults near the Henni salt pillow exhibit a maximum slip tendency of 0.33. NW/SE and NNW/SSE-oriented faults show higher slip tendency, in accordance with the azimuth of maximum horizontal stress in the area \citep{Ahlers2022}. The critical dip and rake are identified as $\approx$55$^{\circ}$ and $\approx$90$^{\circ}$, respectively. Assuming a continuous rupture area along each layer, it is estimated that the maximum potential moment magnitudes are $\approx$2 $M_W$. When the height of the rupture plane is limited two times the height of the reservoir unit, the maximum potential moment magnitude is 1.39 $M_W$. For the majority of the faults, the local maxima of the slip tendency does not correlate with the local maxima of potential moment magnitude. However, several faults in the upper part of the Mesozoic near Henni South show relatively high slip tendency and above-average potential moment magnitude. Although these faults are located above the main seal, there remains a potential reactivation due to the potential leakage pathway through the fault zone. {When analyzed considering the stress tensors according to \cite{Ahlers2022}, higher slip tendencies are obtained, suggesting that the faults are critically stressed and prone to reactivation. These results highlight the importance of a further field investigation to confirm the present-day stress regime as well as the mechanical and rheological properties of the faults. }

\bmhead{Acknowledgements}

The first author declares financial support received for the research, authorship and publication of this article. The study was funded by the Federal Ministry of Education and Research of Germany (BMBF) in the framework of ‘‘GEOSTOR’’ (grant number 03F0893C and 03F0963B), one of the six research consortia of the German Marine Research Alliance (DAM) research mission "Marine carbon sinks in decarboniztion pathways" (CDRmare). 
We also thank AspenTech for providing GOCAD/SKUA/EPOS Software Package licenses via the Academic Software Programme (https://www.aspentech. com/en/academicprogram) to support BGR as a national geological service in non-profit work for the public and education.
The authors would like to also express our gratitude to Firdovsi Gazansade and Sebastian Bauer, whose insight and help have been instrumental for this study.


\bibliography{2_ReferencesClean}

\begin{thebibliography}{59}
\providecommand{\natexlab}[1]{#1}
\providecommand{\url}[1]{{#1}}
\providecommand{\urlprefix}{URL }
\providecommand{\doi}[1]{\url{https://doi.org/#1}}
\providecommand{\eprint}[2][]{\url{#2}}
 \bibcommenthead

\bibitem[{Ahlers et~al(2022)Ahlers, R\"ockel, Hergert, Reiter, Heidbach, Henk,
  M\"uller, Morawietz, Scheck-Wenderoth, and Anikiev}]{Ahlers2022}
Ahlers S, R\"ockel L, Hergert T, et~al (2022) {The crustal stress field of
  Germany: a refined prediction}. Geothermal Energy 10(10).
  \doi{10.1186/s40517-022-00222-6}

\bibitem[{Amos et~al(2014)Amos, Audet, Hammond, Bürgmann, Johanson, and
  Blewitt}]{Amos2014}
Amos CB, Audet P, Hammond WC, et~al (2014) {Uplift and seismicity driven by
  groundwater depletion in central California}. Nature 509:483--6.
  \doi{10.1038/nature13275}

\bibitem[{Baldschuhn et~al(1994)Baldschuhn, Best, Binot, Deneke, Frisch,
  Jürgens, Kockel, Röhling, Sattler-Kosinowski, Schmitz, Stancu-Kristoff, and
  Zirngast}]{Baldschuhn1994}
Baldschuhn R, Best G, Binot F, et~al (1994) {Geotektonischer Atlas
  NW-Deutschland 1:300 000 – 1. Subcrop map of the Base Lower Cretaceous
  unconfomity: 5 S.} Tech. rep., {Bundesanstalt für Geowissenschaften und
  Rohstoffe (BGR), Hannover}

\bibitem[{Bense and Jähne-Klingberg(2017)}]{BENSE2017}
Bense F, Jähne-Klingberg F (2017) {Storage potentials in the deeper subsurface
  of the Central German North Sea}. Energy Procedia 114:4595--4622.
  \doi{https://doi.org/10.1016/j.egypro.2017.03.1580},
  \urlprefix\url{https://www.sciencedirect.com/science/article/pii/S1876610217317745},
  13th International Conference on Greenhouse Gas Control Technologies,
  GHGT-13, 14-18 November 2016, Lausanne, Switzerland

\bibitem[{Bense et~al(2022)Bense, Deutschmann, Dzieran, Hese, Höding, Jahnke,
  Lademann, Liebsch-Dörschner, Müller, Obst, Offermann, Schilling, and
  Wächter}]{Bense2022}
Bense F, Deutschmann A, Dzieran L, et~al (2022) {Potenziale des unterirdischen
  Speicher- und Wirtschaftsraumes im Norddeutschen Becken (TUNB) - Phase 2:
  Parametrisierung. – Abschlussbericht}. Tech. rep., {Bundesanstalt für
  Geowissenschaften und Rohstoffe (BGR), Hannover}

\bibitem[{Brückner-Röhling et~al(2005)Brückner-Röhling, Forsbach, and
  Kockel}]{Bruekner2005}
Brückner-Röhling S, Forsbach H, Kockel F (2005) {The structural development
  of the German North Sea sector during the Tertiary and the Quaternary}.
  Zeitschrift der Deutschen Gesellschaft für Geowissenschaften 156:341--355

\bibitem[{Cheng et~al(2023)Cheng, Liu, Xu, Zhang, Zhang, Xing, Feng, and
  Xia}]{CHENG2023}
Cheng Y, Liu W, Xu T, et~al (2023) {Seismicity induced by geological
  CO\textsubscript{2} storage: A review}. Earth-Science Reviews 239:104369.
  \doi{https://doi.org/10.1016/j.earscirev.2023.104369}

\bibitem[{Clarke et~al(2014)Clarke, Eisner, Styles, and Turner}]{Clarke2014}
Clarke H, Eisner L, Styles P, et~al (2014) {Felt seismicity associated with
  shale gas hydraulic fracturing: The first documented example in Europe}.
  Geophysical Research Letters 41(23):8308--8314.
  \doi{https://doi.org/10.1002/2014GL062047}

\bibitem[{Collettini and Trippetta(2007)}]{COLLETTINI2007}
Collettini C, Trippetta F (2007) A slip tendency analysis to test mechanical
  and structural control on aftershock rupture planes. Earth and Planetary
  Science Letters 255(3):402--413.
  \doi{https://doi.org/10.1016/j.epsl.2007.01.001}

\bibitem[{Diehl et~al(2017)Diehl, Kraft, Kissling, and Wiemer}]{Diehl2017}
Diehl T, Kraft T, Kissling E, et~al (2017) {The induced earthquake sequence
  related to the St. Gallen deep geothermal project (Switzerland): Fault
  reactivation and fluid interactions imaged by microseismicity}. Journal of
  Geophysical Research: Solid Earth 122(9):7272--7290.
  \doi{https://doi.org/10.1002/2017JB014473}

\bibitem[{Ehrhardt et~al(2021)Ehrhardt, Barckhausen, Behrens, Demir, Ebert,
  Engels, Hahn, Kuhlamm, Schnabel, Steinborn, and Stück}]{Ehrhardt2021}
Ehrhardt A, Barckhausen U, Behrens T, et~al (2021) {High Resolution Reflection
  Seismic Imaging of the Cenozoic Barrier Structures of the West-Schleswig
  Block and the Fluid Migration System of the blowout structure 'Figge Maar',
  Cruise Report No. MSM 97 (GPF 20-3\_085), 13.11.2020-25.11.2020, Emden
  (Germany) -- Emden (Germany)}. Tech. rep., {Bundesanstalt für
  Geowissenschaften und Rohstoffe (BGR), Hannover},
  \doi{https://doi.org/10.48433/cr_msm97}

\bibitem[{Ellsworth(2013)}]{Ellsworth2013}
Ellsworth WL (2013) Injection-induced earthquakes. Science 341(6142):1225942.
  \doi{10.1126/science.1225942}

\bibitem[{Fellgett et~al(2018)Fellgett, Kingdon, Williams, and
  Gent}]{FELLGETT2018}
Fellgett MW, Kingdon A, Williams JD, et~al (2018) {Stress magnitudes across UK
  regions: New analysis and legacy data across potentially prospective
  unconventional resource areas}. Marine and Petroleum Geology 97:24--31.
  \doi{https://doi.org/10.1016/j.marpetgeo.2018.06.016}

\bibitem[{Frohlich et~al(2016)Frohlich, DeShon, Stump, Hayward, Hornbach, and
  Walter}]{Frohlich2016}
Frohlich C, DeShon H, Stump B, et~al (2016) {A historical review of induced
  earthquakes in Texas}. Seismological Research Letters 87(4):1022--1038.
  \doi{10.1785/0220160016}

\bibitem[{Fuhrmann et~al(2024)Fuhrmann, Knopf, Thöle, Kästner, Ahlrichs,
  Stück, Schlieder-Kowitz, and Kuhlmann}]{FUHRMANN2024}
Fuhrmann A, Knopf S, Thöle H, et~al (2024) {CO\textsubscript{2} storage
  potential of the Middle Buntsandstein Subgroup - German sector of the North
  Sea}. International Journal of Greenhouse Gas Control 136:104175.
  \doi{https://doi.org/10.1016/j.ijggc.2024.104175}

\bibitem[{Gan and Frohlich(2013)}]{Gan2013}
Gan W, Frohlich C (2013) {Gas injection may have triggered earthquakes in the
  Cogdell oil field, Texas}. Proceedings of the National Academy of Sciences
  110(47):18786--18791. \doi{10.1073/pnas.1311316110}

\bibitem[{Giardini(2009)}]{Giardini2009}
Giardini D (2009) Geothermal quake risks must be faced. Nature 462:848--9.
  \doi{10.1038/462848a}

\bibitem[{Gibowicz(2009)}]{GIBOWICZ2009}
Gibowicz SJ (2009) {Chapter 1 - Seismicity Induced by Mining: Recent Research}.
  In: Advances in Geophysics, Advances in Geophysics, vol~51. Elsevier, p
  1--53, \doi{https://doi.org/10.1016/S0065-2687(09)05106-1}

\bibitem[{Günter(2011)}]{Leydecker2011}
Günter L (2011) Erdbebenkatalog für deutschland mit randgebieten für die
  jahre 800 - 2008 (earthquake catalogue for germany and adjacent areas for the
  years 800 to 2008). Geologisches Jahrbuch E 59:1 -- 198

\bibitem[{Goertz-Allmann et~al(2024)Goertz-Allmann, Langet, Iranpour, Kühn,
  Baird, Oates, Rowe, Harvey, Oye, and Nakstad}]{GOERTZALLMANN2024}
Goertz-Allmann BP, Langet N, Iranpour K, et~al (2024) {Effective microseismic
  monitoring of the Quest CCS site, Alberta, Canada}. International Journal of
  Greenhouse Gas Control 133:104100.
  \doi{https://doi.org/10.1016/j.ijggc.2024.104100}

\bibitem[{Grigoli et~al(2017)Grigoli, Cesca, Priolo, Rinaldi, Clinton, Stabile,
  Dost, Fernandez, Wiemer, and Dahm}]{Grigoli2017}
Grigoli F, Cesca S, Priolo E, et~al (2017) {Current challenges in monitoring,
  discrimination, and management of induced seismicity related to underground
  industrial activities: A European perspective}. Reviews of Geophysics
  55(2):310--340. \doi{https://doi.org/10.1002/2016RG000542}

\bibitem[{Grünthal and Wahlström(2003)}]{Grunthal2003}
Grünthal G, Wahlström R (2003) {An earthquake catalogue for central, northern
  and northwestern Europe based on M\textsubscript{W} magnitudes}. Scientific
  Technical Report STR 03:143. \doi{https://doi.org/10.2312/GFZ.b103-030104}

\bibitem[{Groß(1986)}]{Gross1986}
Groß U (1986) {Gaspotential Deutsche Nordsee – Die regionale Verteilung der
  seismischen Anfangsgeschwindigkeiten in der Deutschen Nordsee}. Tech. rep.,
  {Bundesanstalt für Geowissenschaften und Rohstoffe (BGR)}, Hannover

\bibitem[{Guha(2001)}]{Guha2001}
Guha S (2001) {Induced Earthquakes}. Springer Dordrecht,
  \doi{https://doi.org/10.1007/978-94-015-9452-3}

\bibitem[{Heidbach et~al(2018)Heidbach, Rajabi, Cui, Fuchs, Müller, Reinecker,
  Reiter, Tingay, Wenzel, Xie, Ziegler, Zoback, and Zoback}]{HEIDBACH2018}
Heidbach O, Rajabi M, Cui X, et~al (2018) {The World Stress Map database
  release 2016: Crustal stress pattern across scales}. Tectonophysics
  744:484--498. \doi{https://doi.org/10.1016/j.tecto.2018.07.007}

\bibitem[{Hough(2014)}]{Hough2014}
Hough SE (2014) Earthquake intensity distributions: A new view. Bulletin of
  Earthquake Engineering 12:135--55. \doi{10.1007/s10518-013-9573-x}

\bibitem[{IPCC(2005)}]{IPCC2005}
IPCC (2005) {Special Report on Carbon Dioxide Capture and Storage }. Report,
  Intergovernmental Panel on Climate Change,
  \urlprefix\url{https://www.ipcc.ch/report/carbon-dioxide-capture-and-storage/}

\bibitem[{Jackson and Hudec(2017)}]{Jackson2017}
Jackson M, Hudec M (2017) Salt tectonics: Principles and practice. Cambridge
  University Press, \doi{https://doi.org/10.1017/9781139003988}

\bibitem[{Jaeger et~al(2011)Jaeger, Cook, and Zimmerman}]{Jaeger2011}
Jaeger J, Cook N, Zimmerman R (2011) Fundamental of rock mechanics, 4 edn.
  Blackwell Publishing, Malden, MA

\bibitem[{Jaritz et~al(1991)Jaritz, Best, Hildebrand, and
  Juergens}]{Jaritz1991}
Jaritz W, Best G, Hildebrand G, et~al (1991) {Regionale Analyse der seismischen
  Geschwindigkeiten in Nordwestdeutschland}. {Geologisches Jahrbuch} Reihe E:
  45:23--57

\bibitem[{Jähne-Klingberg et~al(2014)Jähne-Klingberg, Wolf, Steuer, Bense,
  Kaufmann, and Weitkamp}]{Jaehne2014}
Jähne-Klingberg F, Wolf M, Steuer S, et~al (2014) {Speicherpotenziale Deutsche
  Nordsee}. Geopotenzial Deutsche Nordsee 110

\bibitem[{Kanamori and Brodsky(2004)}]{Kanamori2004}
Kanamori H, Brodsky E (2004) The physics of earthquakes. Physics Today 54.
  \doi{10.1088/0034-4885/67/8/R03}

\bibitem[{Keranen and Weingarten(2018)}]{Keranen2018}
Keranen KM, Weingarten M (2018) Induced seismicity. Annual Review of Earth and
  Planetary Sciences 46:149--174.
  \doi{https://doi.org/10.1146/annurev-earth-082517-010054}

\bibitem[{Keranen et~al(2014)Keranen, Weingarten, Abers, Bekins, and
  Ge}]{Keranen2014}
Keranen KM, Weingarten M, Abers GA, et~al (2014) {Sharp increase in central
  Oklahoma seismicity since 2008 induced by massive wastewater injection}.
  Science 345(6195):448--451. \doi{10.1126/science.1255802}

\bibitem[{Kockel(1995)}]{Kockel1995}
Kockel F (ed)  (1995) Structural and Palaeogeographical Development of the
  German North Sea Sector. Schweizerbart Science Publishers, Stuttgart,
  Germany,
  \urlprefix\url{http://www.schweizerbart.de//publications/detail/isbn/9783443110260/Kockel\_Structural\_and\_Palaeogeographic},
  with contribution from R. Baldschuhn, G. Best, F. Binot, U. Frisch, U. Gross,
  U. Jürgens, H.-G. Röhling \& S. Sattler-Kosinowski, S.

\bibitem[{Luu et~al(2022)Luu, Schoenball, Oldenburg, and Rutqvist}]{Luu2022}
Luu K, Schoenball M, Oldenburg CM, et~al (2022) {Coupled Hydromechanical
  Modeling of Induced Seismicity From CO\textsubscript{2} Injection in the
  Illinois Basin}. Journal of Geophysical Research: Solid Earth
  127(5):e2021JB023496. \doi{https://doi.org/10.1029/2021JB023496}

\bibitem[{Moeck et~al(2009)Moeck, Kwiatek, and Zimmermann}]{MOECK2009}
Moeck I, Kwiatek G, Zimmermann G (2009) Slip tendency analysis, fault
  reactivation potential and induced seismicity in a deep geothermal reservoir.
  Journal of Structural Geology 31(10):1174--1182.
  \doi{https://doi.org/10.1016/j.jsg.2009.06.012}

\bibitem[{Morawietz et~al(2020)Morawietz, Heidbach, Reiter, Ziegler, Rajabi,
  Zimmermann, M\"uller, and Tingay}]{Morawietz2020}
Morawietz S, Heidbach O, Reiter K, et~al (2020) {An open-access stress
  magnitude database for Germany and adjacent regions}. Geothermal Energy
  8(25):1777--1799. \doi{10.1186/s40517-020-00178-5}

\bibitem[{Morris et~al(1996)Morris, Ferrill, and Henderson}]{Morris1996}
Morris A, Ferrill DA, Henderson D (1996) {Slip-tendency analysis and fault
  reactivation}. Geology 24(3):275--278.
  \doi{10.1130/0091-7613(1996)024<0275:STAAFR>2.3.CO;2}

\bibitem[{{Natl. Res. Counc.}(2013)}]{NAP13355}
{Natl. Res. Counc.} (2013) Induced Seismicity Potential in Energy Technologies.
  The National Academies Press, Washington, DC, \doi{10.17226/13355}

\bibitem[{Noy et~al(2012)Noy, Holloway, Chadwick, Williams, Hannis, and
  Lahann}]{NOY2012}
Noy D, Holloway S, Chadwick R, et~al (2012) {Modelling large-scale carbon
  dioxide injection into the Bunter Sandstone in the UK Southern North Sea}.
  International Journal of Greenhouse Gas Control 9:220--233.
  \doi{https://doi.org/10.1016/j.ijggc.2012.03.011}

\bibitem[{Orlic(2016)}]{ORLIC2016}
Orlic B (2016) {Geomechanical effects of CO\textsubscript{2} storage in
  depleted gas reservoirs in the Netherlands: Inferences from feasibility
  studies and comparison with aquifer storage}. Journal of Rock Mechanics and
  Geotechnical Engineering 8(6):846--859.
  \doi{https://doi.org/10.1016/j.jrmge.2016.07.003}

\bibitem[{Peters(2007)}]{Peters2007}
Peters G (2007) {Active tectonivs in the Upper Rhine Graben: Integration of
  paleoseismology, geomorphology and geomechanical modeling}. PhD thesis,
  {Amsterdam, Vrije Univ.}

\bibitem[{Reyer and Philipp(2014)}]{Reyer2014}
Reyer D, Philipp SL (2014) Empirical relations of rock properties of outcrop
  and core samples from the {Northwest German Basin} for geothermal drilling.
  Geothermal Energy Science 2(1):21--37.
  \urlprefix\url{http://resolver.sub.uni-goettingen.de/purl?gldocs-11858/6861}

\bibitem[{R\"ockel et~al(2022)R\"ockel, Ahlers, M\"uller, Reiter, Heidbach,
  Henk, Hergert, and Schilling}]{Roeckel2022}
R\"ockel L, Ahlers S, M\"uller B, et~al (2022) {The analysis of slip tendency
  of major tectonic faults in Germany}. Solid Earth 13(6):1087--1105.
  \doi{10.5194/se-13-1087-2022}

\bibitem[{Rutqvist et~al(2016)Rutqvist, Rinaldi, Cappa, Jeanne, Mazzoldi, Urpi,
  Guglielmi, and Vilarrasa}]{Rutqvist2016}
Rutqvist J, Rinaldi AP, Cappa F, et~al (2016) {Fault activation and induced
  seismicity in geological carbon storage – Lessons learned from recent
  modeling studies}. Journal of Rock Mechanics and Geotechnical Engineering
  8(6):789--804. \doi{https://doi.org/10.1016/j.jrmge.2016.09.001}

\bibitem[{Samuelson and Spiers(2012)}]{SAMUELSON2012}
Samuelson J, Spiers CJ (2012) {Fault friction and slip stability not affected
  by CO\textsubscript{2} storage: Evidence from short-term laboratory
  experiments on North Sea reservoir sandstones and caprocks}. International
  Journal of Greenhouse Gas Control 11:S78--S90.
  \doi{https://doi.org/10.1016/j.ijggc.2012.09.018}, cATO: CCS Research in the
  Netherlands

\bibitem[{Scholz(1998)}]{Scholz1998}
Scholz CH (1998) Earthquakes and friction laws. Nature 391:37 -- 42.
  \doi{https://doi.org/10.1038/34097}

\bibitem[{Scholz(2019)}]{Scholz_2019}
Scholz CH (2019) The Mechanics of Earthquakes and Faulting, 3rd edn. Cambridge
  University Press, \doi{https://doi.org/10.1017/9781316681473}

\bibitem[{Stork et~al(2018)Stork, Nixon, Hawkes, Birnie, White, Schmitt, and
  Roberts}]{STORK2018}
Stork A, Nixon C, Hawkes C, et~al (2018) {Is CO\textsubscript{2} injection at
  Aquistore aseismic? A combined seismological and geomechanical study of early
  injection operations}. International Journal of Greenhouse Gas Control
  75:107--124. \doi{https://doi.org/10.1016/j.ijggc.2018.05.016}

\bibitem[{Thöle et~al(2014)Thöle, Gaedicke, Kuhlmann, and
  Reinhard}]{Thoele2014}
Thöle H, Gaedicke C, Kuhlmann G, et~al (2014) {Late Cenozoic sedimentary
  evolution of the German North Sea – A seismic stratigraphic approach}.
  Newsletters on Stratigraphy 47:299--329

\bibitem[{Trifu(2002)}]{Trifu2002}
Trifu C (2002) {The Mechanism of Induced Seismicity}. Pageoph Topical Volumes,
  Birkhäuser Basel, \doi{https://doi.org/10.1007/978-3-0348-8179-1}

\bibitem[{Vadacca et~al(2021)Vadacca, Rossi, Scotti, and
  Buttinelli}]{Vadacca2021}
Vadacca L, Rossi D, Scotti A, et~al (2021) {Slip Tendency Analysis, Fault
  Reactivation Potential and Induced Seismicity in the Val d'Agri Oilfield
  (Italy)}. Journal of Geophysical Research: Solid Earth 126(1):2019JB019185.
  \doi{https://doi.org/10.1029/2019JB019185}

\bibitem[{Verdon and Stork(2016)}]{VERDON2016}
Verdon J, Stork A (2016) Carbon capture and storage, geomechanics and induced
  seismic activity. Journal of Rock Mechanics and Geotechnical Engineering
  8(6):928--935. \doi{https://doi.org/10.1016/j.jrmge.2016.06.004}

\bibitem[{Verdon et~al(2011)Verdon, Kendall, White, and Angus}]{VERDON2011}
Verdon J, Kendall JM, White D, et~al (2011) {Linking microseismic event
  observations with geomechanical models to minimise the risks of storing
  CO\textsubscript{2} in geological formations}. Earth and Planetary Science
  Letters 305(1):143--152. \doi{https://doi.org/10.1016/j.epsl.2011.02.048}

\bibitem[{Wallmann(2023)}]{Wallman2023}
Wallmann K (2023) {CCS (Carbon Capture and Storage):
  CO\textsubscript{2}-Speicherung unter der Nordsee}. In: L. LJ, Graßl H,
  Breckle S, et~al (eds) Warnsignal-Klima, pp 120--125,
  \doi{http://doi.org/10.25592/uhhfdm.12816}

\bibitem[{Weemstra et~al(2022)Weemstra, Kettlety, Kühn, Martuganova,
  Schweitzer, Baptie, and Dahl-Jensen}]{SHARP2022}
Weemstra C, Kettlety T, Kühn D, et~al (2022) Integrated earthquake locations
  and magnitudes plus focal mechanisms for the north sea \& construction of a
  velocity model. Tech. rep., Delft University of Technology, NORSAR, GEUS,
  University of Oxford, BGS, Equinor, Shell, BP,
  \urlprefix\url{https://www.sharp-storage-act.eu/publications--results/}

\bibitem[{White et~al(2002)White, Traugott, and Swarbrick}]{White2002}
White A, Traugott M, Swarbrick R (2002) The use of leak-off tests as means of
  predicting minimum in-situ stress. Petroleum Geoscience 8:189--193.
  \doi{10.1144/petgeo.8.2.189}

\bibitem[{White and Foxall(2016)}]{White2016}
White J, Foxall W (2016) {Assessing induced seismicity risk at
  CO\textsubscript{2} storage projects: Recent progress and remaining
  challenges}. International Journal of Greenhouse Gas Control 49:413--424.
  \doi{https://doi.org/10.1016/j.ijggc.2016.03.021}

\end{thebibliography}

\end{document}